\documentstyle[preprint,prb,aps,psfig,epsf,floats]{revtex}
\newcommand{\beq}{\begin{equation}}
\newcommand{\eeq}{\end{equation}}

\newcommand{\bfi}{\begin{figure}}
\newcommand{\efi}{\end{figure}}

\begin{document}
\title{\bf Holstein model in infinite dimensions at half-filling}

\vspace{3cm}
\author{{Patrizia Benedetti} and {Roland Zeyher}}

\address{Max-Planck-Institut\  f\"ur\
Festk\"orperforschung,\\ Heisenbergstr.1, 70569 Stuttgart, Germany \\}

\date{\today }

\vspace{3cm}

\maketitle

\begin{abstract}

The normal state of the Holstein model is studied at half-filling in 
infinite dimensions and in the adiabatic regime. The dynamical mean-field
equations are solved using perturbation expansions around the extremal
paths of the effective action for the atoms. We find that the 
Migdal-Eliashberg expansion breaks down in the metallic state 
if the electron-phonon coupling $\lambda$ exceeds a value of about 1.3
in spite of the fact that the formal expansion parameter
$\lambda \omega_0/E_F$ ($\omega_0$ is the phonon frequency, $E_F$ the
Fermi energy) is much smaller than 1. The breakdown is due to
the appearance of more than one extremal path of the action.
We present numerical results which illustrate in detail the evolution of
the local Green's function, the self-energy and the effective atomic
potential as a function of $\lambda$.

\end{abstract}

\newpage 

\section{Introduction}

Studies on the microscopic mechanism of high-T$_c$ superconductivity
and of the colossal magnetoresistance have shown the need for a more
detailed investigation of simple models for the electron-phonon coupling.
One generic model is the Holstein model \cite{Holstein} which consists 
of conduction electrons
interacting locally with optical phonons. The case of just one electron
has received a lot of attention in the past. In particular, the
formation of small and large polarons and their motions have been
studied considering both the adiabatic and the anti-adiabatic cases
\cite{Holstein,Emin,Kabanov,Fehske1,Ciuchi1,Ciuchi2}. 
However, the application of the obtained results to high-T$_c$
superconductors \cite{Polarons} or to the manganites\cite{Min} is problematic. 
In these
systems the Fermi energy $E_F$ is substantially larger than typical phonon 
energies $\omega_0$. One thus deals in these systems with a large
Fermi surface where the Migdal-Eliashberg \cite{Migdal-Elia}
self-consistent perturbation
theory, governed by the small parameter $\lambda \omega_0 /E_F$
($\lambda$ is the electron-phonon coupling constant), should be more 
appropriate than the use of canonical transformations
typical for the one-electron problem.

 From the studies on the single polaron problem it is known that polaron
formation, i.e., the sudden and large mass enhancement of the electron,
takes place around $\lambda \sim 1$. Associated with this is a breakdown of
simple perturbation theory due to the appearance of a bound state in
the electron-phonon system. Whether something similar happens
in the case of many electrons and a large Fermi surface is still a matter
of controversy. On the one hand, it has been argued that, due
to the Migdal theorem, self-consistent theory works even if $\lambda$ is not 
small. It is claimed that the only condition for the validity of the 
self-consistent 
perturbation theory is that $\lambda \omega_0 / E_F \ll 1$ 
holds \cite{Migdal-Elia}.
As a result many treatments in the past, for instance, for 
high-T$_c$ superconductors, used values for $\lambda$ which are much 
larger than 1 \cite{Reeves}. On the other
hand, it has been claimed that Migdal-Eliashberg perturbation theory
breaks down if $\lambda$ exceeds
a critical value somewhere between 1 and 2, irrespective of the 
electronic density and the size of the Fermi surface \cite{Alex-Mott}.
The arguments for this, however,
were based on the Lang-Firsov transformation \cite{Lang-Firsov} which is more appropriate
in the anti-adiabatic than in the adiabatic limit and they seem to be supported
by numerical calculations \cite{Fehske2}. One aim of our
contribution is to study the possible breakdown of Migdal-Eliashberg
perturbation theory and polaron formation in the adiabatic regime
where the hopping and the Fermi energy are much larger than the phonon
energies. Another aim is the calculation of phonon renormalization effects
caused by an arbitrarily strong electron-phonon coupling.

A suitable method to study the above problem is the dynamical mean-field
theory \cite{Kotliar,Metzner}. This method becomes exact in the limit of infinite spatial
dimensions. It allows to map the lattice problem onto Anderson's
single impurity problem embedded in an effective bath created by all
the other sites and interacting with the electrons on the impurity 
in a dynamical, time-dependent way. In spite of the substantial
simplifications introduced by infinite dimensions the resulting 
equations are in general still too complicated to admit
exact analytic solutions. For the Holstein model some results from
Monte Carlo simulations are available \cite{Freer1,Freer2,Freer3,Freer4,Freer5}. 
Various approximate treatments
have also been performed such as iterated perturbation theory \cite{Freer4},
self-consistent perturbation theory \cite{Freer3} and the semiclassical 
approximation \cite{Millis1,Millis2}.

In the following we will formulate the dynamical mean-field equations
in terms of path integrals. The phonon coordinate will not be
treated as a classical variable without dynamics as in the semiclassical
approach. Instead, we expand the path integrals around the paths
which extremize the action. For small $\lambda$ there is only one 
extremal and trivial path and Migdal-Eliashberg expansion is valid.
Beyond a critical value for $\lambda$ more than one extremal paths
exist signalizing the breakdown of Migdal-Eliashberg theory
though the system is still in the metallic state. Our treatment
can be applied in the adiabatic regime to all coupling strengths
 $\lambda$ and contains
correctly the weak- and the strong-coupling cases. The paper is
organized as follows: In section II the model and the dynamical
mean-field equations are formulated. In section III the electrons
are approximately integrated out in the path integral using a
gradient expansion in time. The resulting effective action for the atoms 
is discussed. Approximate expressions for the electron Green's function
are obtained in section IV. Section V contains numerical results for a 
Bethe lattice both for the atomic and the electronic properties.
This section illustrates in detail the evolution of basic quantities of
the dynamical mean field theory such as the local Green's function,
the electronic self-energy, the effective ``unperturbed'' Green's function
and the effective atomic potential as a function of $\lambda$.
Finally, section VI contains the conclusions.

\section{Models and Dynamical Mean-Field equations}

The Hamiltonian for the Holstein model can be written as\cite{Holstein}

\begin{equation}
H =-\sum_{ij\sigma}t_{ij} (c^{+}_{i \sigma}c_{j\sigma}+h.c.)
+ \frac{1}{2} \sum_{i}
( (\partial_{t}\phi_{i})^{2}+\omega_{0}^{2}\phi_{i}^{2})+
g\sum_{i \sigma} n_{i \sigma}\phi_{i}.
\end{equation}

It describes a system of electrons interacting locally with dispersionless 
phonons. $i,j$ denote the sites of a lattice,
$\omega_{0}$ is the frequency of the phonons, $t_{ij}$ is the hopping 
matrix element of the electrons, and $g$ the coupling constant between 
electrons and phonons. $\phi_i$ is the displacement operator of the
atom at site $i$ which can be expressed in terms of bosonic
creation and annihilation operators $a_i^\dagger,a_i$ as
\begin{equation}
\phi_{i}=\frac{1}{\sqrt{2\omega_{0}}}(a_{i}+a^{+}_{i}).
\end{equation}
$c_{i,\sigma}^\dagger,c_{i,\sigma}$ are creation and annihilation operators 
for electrons at site $i$ and spin direction $\sigma$ and 
$n_{i\sigma}= c^{\dagger}_{i \sigma}c_{i\sigma}$ is the density operator
of electrons.
The partition function of the model can be written as a path integral 
over anticommutating Grassmann
numbers $c^\ast,c$ (replacing the electronic operators),  and phonon fields,
 
\begin{equation}
Z=\int{\it D}\phi{\it D}c^{*}{\it D}c\,\, e^{{S[c^{*},c,\phi]}},
\end{equation}
where S is the action associated with the  Hamiltonian (1). 
\par

 In the limit of infinite spatial dimensions, $d \rightarrow \infty$,
the dynamical mean-field approximation for $H$ becomes exact. As a
result the many-body problem Eq.(1) can be mapped onto an impurity model
embedded in a dynamical bath \cite{Kotliar}. Denoting the one-electron Green's
function associated with the Hamiltonian of Eq.(1) by $G({
\bf k},i\omega_n)$, the self-energy $\Sigma$ is defined via Dyson's
equation
\begin{equation}
G^{-1}({\bf k},i\omega_n) = G^{(0)^{-1}}({\bf k},i\omega_n) - \Sigma(i
\omega_n).
\end{equation}
$G^{(0)}({\bf k},i\omega_n)$ is the free electron Green's function associated with the 
hopping term in Eq.(1). Using the Matsubara formalism $\omega_n$
are fermionic Matsubara frequencies $\omega_n = (2n+1)\pi T$
where n is an integer and $T$ the temperature. A very important 
simplification in infinite dimensions results from the fact that the
self-energy $\Sigma$ is independent of the momentum $\bf k$ and
depends only on the frequency \cite{Metzner}. The self-energy $\Sigma$ also represents
the self-energy of a site-independent impurity problem with the effective 
action $S_{eff}$ given by
\begin {eqnarray}
S_{eff} &=&\int^{\beta}_{0}d\tau \int^{\beta}_{0}d\tau'\sum_{\sigma}c_{\sigma}^{*}(\tau)G_{0}^{-1}(\tau-\tau')c_{\sigma}(\tau')
\\
&-&\frac{1}{2} \int^{\beta}_{0}d\tau\,\phi(\tau)(-\frac{d^{2}}{d\tau^{2}}+\omega_{0}^{2})\phi(\tau) 
-g\int^{\beta}_{0}d\tau\sum_{\sigma} n_{\sigma}(\tau)\phi(\tau). \ \nonumber
\end{eqnarray}

$G_0$ describes a dynamical effective field acting on the impurity
electrons due to the presence of all the other electrons in the crystal.
It plays the role of a bare Green's function in $S_{eff}$ and contains all
the information on the influence of the surroundings on the impurity
electrons. $\beta$ in Eq.(5) stands for the inverse temperature $1/T$.
Denoting the exact one-electron Green's function associated
with $S_{eff}$ by $G_{loc}$ Dyson's equation of the impurity problem reads
\begin{equation}
G^{-1}_{loc}(i\omega_n) = G_0^{-1}(i\omega_n) - \Sigma(i\omega_n).
\end{equation}
Finally, the so far unknown function $G_0$ is determined by the
condition that $G_{loc}$ coincides with the local Green's function
calculated from the lattice Green's function $G({\bf k},i\omega_n)$, i.e.,
\begin{equation}
G_{loc}(i\omega_{n})=\int d\epsilon\, \frac{N(\epsilon)}{i\omega_{n}
-\epsilon+\mu_{0}-\Sigma(i\omega_{n})}.
\end{equation}
$\mu_0$ is the bare chemical potential and $N(\epsilon)$ is the bare 
density of electron states associated with
the original hopping term in Eq.(1). In the following we will mainly
consider the semicircular density of states of the Bethe lattice \cite{Economu} given by
\begin{equation}
N(\epsilon)=\frac{\sqrt{(4t^{2}-\epsilon^{2})}}{2\pi t^{2}}.
\end{equation}
This lattice has an infinite coordination number $z \rightarrow \infty$
and a rescaled nearest-neighbor 
hopping term $t_{ij}=t/ \sqrt{z}$\,. In this particular case the relation 
between $G_{0}$ and $G_{loc}$ becomes
\begin{equation}
G_{0}^{-1}(i\omega_{n})= i\omega_{n}+\mu_{0}-t^{2}G_{loc}(i\omega_{n}).
\end{equation}
Another popular density of states function is that of the Lorentz model
\begin{equation}
N(\epsilon) = {t \over{\pi(\epsilon^2+t^2)}}.
\end{equation}
It describes a model with infinite range hopping terms and
a function $G_0$ which simply is given by

\begin{equation} 
G_0(i\omega_n) = i \omega_n + \mu_0 +it sgn\omega_n.
\end{equation}
$G_0$ decouples in this case from all the other quantities.
This model is therefore especially simple
allowing often analytic expressions which capture essential features 
of models with more realistic density of states functions. On the other 
hand it misses the feedback from the other quantities and thus is
unable to describe, for instance, the Mott-Hubbard transition from a metal
to an insulator. \par

Eqs.(5)-(7) represent a closed set of equations. They have been
derived under the assumption that no long-range order is present which 
will be assumed in the following. The above equations can be solved 
for large regions in parameter space numerically using Monte Carlo 
or diagonalization methods \cite{Kotliar}. In the following we will use a more analytic
method which gives directly some insight into the underlying physics.

\section{Effective action for the atoms}

According to Eq.(5) $S_{eff}$ is bilinear in the electronic variables.
This means that the integration over electrons in the partition
function Eq.(3) can be carried out yielding the following action $S_{ph}$
for the phonon field alone:
\begin{equation}
S_{ph}=-\frac{1}{2} \int^{\beta}_{0}d\tau\,\phi(\tau)(-\frac{d^{2}}
{d\tau^{2}}+\omega_{0}^{2})\phi(\tau)+ Tr\log(G_{0}^{-1}-g\phi).
\end{equation}
Here $Tr$ means the trace over electronic variables, i.e., an integration
over $\tau$ from $0$ to $\beta$ and a sum over the two spin components.
The argument of the logarithm in Eq.(12) is a nondiagonal matrix
consisting of the non-diagonal part $G_0^{-1}$
and the diagonal part $\phi$. Similarly, in $\omega_n$-space the
first contribution $G_0^{-1}$ would be diagonal but the second one,
$\phi$ would become non-diagonal. \par
Defining the expectation value of $\phi$ in the usual way by
\begin{equation}
< \phi > = {\int D \phi \phi e^{S_{ph}}}/Z,
\end{equation}
with
\begin{equation}
Z = \int D \phi e^{S_{ph}},
\end{equation}
there is no reason to expect that this static expectation value
is zero even in the half-filled case. Indeed, writing down
the equation of motion of Heisenberg operators using the original
Hamiltonian Eq.(1) and considering the static limit we have
\begin{equation}
< \phi > = -{g \over {\omega_0^2}} < n >.
\end{equation}
$<n>$ is the average number of electrons per site.
It is therefore convenient to split $\phi$
into a static and a fluctuation part according to
\begin{equation}
\phi = < \phi> + \tilde{\phi},
\end{equation}
where $< \phi >$ is defined by the condition $< \tilde{\phi} > = 0$.
This shift can be interpreted as a Hartree term $\Sigma_H = g< \phi >$
in the self-energy of $G_0$. Since this term is frequency independent
it just renormalizes the bare chemical potential as
\begin{equation}
\mu = \mu_0 - \Sigma_H.
\end{equation}
From now on we will assume that this shift has been performed and
that the chemical potential in $G_0$ is $\mu$ and no longer the bare one
$\mu_0$. \par
The main problem left is the evaluation of the $Tr\;log$ term in 
Eq.(12). For this we note that
two independent parameters can be naturally defined in the Holstein 
model: one is the electron-phonon coupling  constant
$\lambda=g^{2}/ \omega_{0}^{2}t$, and the other one is the adiabatic 
parameter  $\gamma= \omega_{0}/t$. 
The adiabatic  regime is defined by  $\gamma \ll 1$, corresponding to the situation
where the electrons move much faster than the phonons. For the calculation of
the $Tr\;log$ term we assume that $ \gamma \ll 1$ holds. $\tilde{\phi}$ varies then slowly
as function of $\tau$ compared with the variations in the electronic
coordinates. An expansion in time gradients  of $\tilde \phi$ represents thus
an expansion in the Migdal parameter $\gamma$ and we can confine ourselves to
the lowest-order  contributions. \par

Expanding the $Tr\;log$ term in powers of $g$ one obtains
\begin{equation}
Tr\; log (G_0^{-1} -g \tilde{\phi}) = Tr\; log G_0^{-1} - \sum_{n=1}^\infty 
{1 \over n} Tr (G_0 g \tilde{\phi})^n.
\end{equation}
The first term on the right-hand side of Eq.(18) is independent of
$\tilde{\phi}$ and thus will be dropped in the following. Writing the
conservation of frequencies explicitly by means of a delta-function
the general term of n-th order in $g$ of the $Tr\; log$  term becomes
\begin{eqnarray}
X_n &=& -2{{g^n}\over n} T \sum_{{\omega_1...\omega_n}\atop \nu}
\int_0^\beta d \tau e^{i(\omega_1+...+\omega_n) \tau} \tilde{\phi}(\omega_1)
G_0(\nu+\omega_1)\cdot \\ \nonumber 
&&\tilde{\phi}(\omega_2) G_0(\nu+\omega_1+\omega_2)...\tilde{\phi}(\omega_n)
G_0(\nu+\omega_1+...+\omega_n).
\end{eqnarray}
The prefactor 2 in Eq.(19) is due to the sum over the two spin directions.
Due to the delta-function there are no restrictions in the sums 
over $\omega_1...\omega_n$ in Eq.(19). For $\gamma \ll 1$ the motion of the atoms
is slow compared to that of the electrons. In leading order in $\gamma$
the electrons interact thus elastically with the phonons, i.e.,
they see the momentary value of the phonon field as a static field.
This means that the phonon frequencies $\omega_1...\omega_n$ 
can be neglected in the electron propagators $G_0$. The sums over
$\omega_1...\omega_n$ can then trivially be carried out in Eq.(19).
Summing also over $n$ and adding the
shifted harmonic potential the total potential $V(\tilde{\phi}(\tau))$
in $S_{ph}$ becomes in this local approximation
\begin{equation}
V(\tilde{\phi}(\tau) )= \frac{\omega_{0}^{2}}{2} (\tilde{\phi}(\tau)
+\langle\phi\rangle )^{2}
-2T \sum_\nu e^{i \nu0{^+}} log(1-g\tilde{\phi}(\tau) G_0(i\nu)). 
\end{equation}
In the last term in Eq.(20) we have added an exponential in order to
regularize the sum over Matsubara frequencies correctly. This regularization
affects only the first-order term in $g$. 
We found that it is not advisable in general 
to carry out the frequency sum in Eq.(20) directly on the computer but
to rewrite it as a contour integral in the complex plane and to evaluate
it as an integral along the cut on the real frequency axis. One  
obtains in this way
\begin{eqnarray}
        V(\tilde{\phi})=\frac{\omega_{0}^{2}}{2} (\tilde{\phi}
+\langle\phi\rangle )^{2} & \nonumber \\ 
-\frac{2}{\pi} \int_{-\infty}^{+\infty}d\omega 
f(\omega) & \left\{ \arctan\left(\frac{g\tilde{\phi}\Im G_{0}
(\omega + i \eta)}
{1-g\tilde{\phi}\Re G_{0}(\omega + i \eta)}\right )-
\pi \theta (g\tilde{\phi} \Re G_{0}(\omega + i \eta)-1)sign(\tilde{\phi})
\right\}.
\end{eqnarray}
Here $f$ denotes the Fermi function, $\Re$ and $\Im$ real and imaginary
parts, respectively, and $\theta$ and $sign$ the theta- and sign-functions,
respectively. In the case of the Lorentz model the integral in Eq.(21)
can be performed analytically and one obtains at zero temperature:

\begin{eqnarray}
V( \tilde{\phi} ) &=& \frac{1}{2} \omega_{0}^{2}(\tilde{\phi}+
\langle\phi\rangle )^{2} -1/ \pi [ -2 g {\tilde \phi} \arctan \left( 
\frac{t}{g \tilde{\phi}} \right)  \nonumber  \\
&+&t \log \left( \frac{t^{2}} { ( g \tilde{\phi} )^{2}+t^{2} } \right)
+\pi (g \tilde{\phi} ) sign(\tilde{\phi})
-\pi g \tilde{\phi}].    
\end{eqnarray}
\par

A conventional way to calculate corrections to the above local
terms is to use a gradient expansion \cite{Muramatsu}.
 One expands then the propagators
in powers of frequencies $\omega_1....\omega_n$ and rewrites these
powers in terms of time derivatives acting on the field $\tilde{\phi}$.
Such a procedure is, however, not possible in our case because
of the appearance of nonanalytic terms. To see this let us first look
at the lowest-order term $n=2$:

\begin{equation}
X_2 = -{g^2 \over 2} Tr({G_0 \tilde{\phi} G_0 \tilde{\phi}}) 
= -{g^2\over 2} \sum_n
\tilde{\phi}(n) \Pi(i \omega_n) \tilde{\phi}(-n),
\end{equation}
with 
\begin{equation}
\Pi(i \omega_n) = 2 \sum_m G_0(i \omega_m) G_0(i \omega_m + i \omega_n).
\end{equation}
Using the spectral representation
\begin{equation}
G_0(i \omega_n) = \int d \epsilon {{\rho(\epsilon)} \over 
{i \omega_n - \epsilon}}
\end{equation}
and carrying out the integrations one obtains 
at small frequencies and low temperatures and also disregarding a constant
term
\begin{equation}
\Pi(i \omega_n) = 2 \pi \rho^2(0) \beta |\omega_n|,
\end{equation}
where $\rho(0)$ is the spectral function at the chemical potential $\mu=0$.
Considering $X_2$ as an self-energy contribution to the unperturbed
phonon propagator in Eq.(12) and performing the analytic continuation
$|\omega_n| \rightarrow \sqrt{-(\omega + i \eta)^2} \rightarrow -i \omega$,
one recognizes that $X_2$ describes the damping of phonons which
is smaller than the phonon frequency $\omega_0$ by the Migdal paramter $\gamma$. 
Eq.(26) shows, however, that such a damping term produces a nonanalytic
dependency in $S_{ph}$ on Matsubara frequencies. 
This rules out a straightforward application of the
usual gradient expansion. Such an expansion would correspond to a 
Taylor expansion of
$\Pi$ in powers of $i \omega_n$. Since $\Pi$ has to be even in
$i \omega_n$ its linear term would vanish and the leading term would
be of second order in $i \omega_n$ in disagreement with the exact result 
Eq.(26).
To overcome this problem we will use in the following a modified
version of a gradient expansion which is able to give the leading
non-local corrections to the local term also in our case. 
\par 
In leading order in $\gamma$ or, equivalently, in the local approximation,
we assumed that the electrons in the general n-th order term $X_n$
given in Eq.(19) interact elastically with the phonons.
The first correction to that obviously is obtained by considering
in this diagram 2 vertices where the electrons are inelastically and
n-2 vertices where the electrons are elastically scattered. The case
of one vertex with inelastic scattering is not possible due to the
conservation of  Matsubara frequencies. The analytic expression for
$X_n$ is, taking the above processes into account, 
\begin{eqnarray}
X_n^{(2)} = -{g^n \over{n}} \sum_{i \neq j} \sum_{{\omega_1...\omega_n}
\atop {\nu}}
T \int_0^\beta d \tau e^{i(\omega_1+...+\omega_n) \tau}
\tilde{\phi}(\omega_1)G_0(i \nu) \tilde{\phi}(\omega_2) G_0(i \nu)... 
\nonumber \\
\tilde{\phi}(\omega_i) G_0(i \nu+\omega_i) \tilde{\phi}(\omega_{i+1})
G_0(i \nu
+ \omega_i)...\tilde{\phi}(\omega_j) G_0(i \nu) \tilde{\phi}(\omega_{j+1})
G_0(i \nu)... \tilde{\phi}(\omega_n) G_0(i \nu).
\end{eqnarray}  
$i$ and $j$ are the two vertices where inelastic scattering occurs. A factor
$1/2$ has also been introduced in order not to overcount the
contributions. However, this factor is compensated by the spin factor 2.
The superscript $2$ at $X$ indicates that all processes
are taken into account where inelastic scattering takes place at two
vertices. \par
The sums over $\omega_1...\omega_i$ and $\omega_{j+1}...\omega_n$
can be carried out in Eq.(27) yielding
\begin{equation}
X_n^{(2)} = -{g^2/{n}} \sum_{i \neq j} \sum_{{\omega_i,\omega_j}\atop
\nu} T \int_0^\beta d \tau e^{i(\omega_i+\omega_j)\tau} \tilde{\phi}(\omega_i)
\tilde{\phi}(\omega_j) \tilde{G}^{(n-j+i-1)}(i \nu) \tilde{G}^{(j-i-1)}.
\end{equation}
$\tilde{G}$ is defined by
\begin{equation}
{\tilde{G}^{-1}}(i \nu) = G_0^{-1}(i \nu) -g \tilde{\phi}(\tau).
\end{equation}
$\tilde{G}^{(i)}$ denotes $\tilde{G}$, calculated in i-th order in the
coupling constant $g$, i.e.,
\begin{equation}
\tilde{G}^{(i)}(i \nu) = (g \tilde{\phi}(\tau))^i G_0^{i+1}(i \nu).
\end{equation}
Keeping only the term $i=1$ in Eq.(28) is equivalent to omitting 
the prefactor $1/n$.
Moreover, the sum over $j \neq i$ can easily be carried out yielding
\begin{equation}
X_n^{(2)} = -g^2 \sum_{\omega_1,\omega_2} \int_0^\beta d \tau
e^{i(\omega_1+\omega_2) \tau} \tilde{\phi}(\omega_1) \tilde{\phi}
(\omega_2) T \sum_\nu
\tilde{G}(i \nu) \tilde{G}(i \nu+i \omega_1),
\end{equation}
where all terms of order $n-2$ have to be taken into account in the
product of the two Green's functions $\tilde{G}$. Performing the sum over
$\nu$ in Eq.(31), using the spectral function $\tilde{\rho}$ of $\tilde{G}$,
and also summing over $\omega_2$ we obtain
\begin{equation}
X_n^{(2)} = -g^2 T \int_0^\beta d \tau d\tau'(\sum_{\omega_1}
e^{i \omega_1(\tau-\tau')} |\omega_1|) \tilde{\phi}(\tau) 
\tilde{\phi}(\tau') F[\phi(\tau)],
\end{equation}
with
\begin{equation}
F[\phi]) = -\pi \tilde{\rho}^2(0).
\end{equation}
$\tilde{\rho}$ is the spectral function of $\tilde G$ which also depends
on $\tilde{\phi}(\tau)$. Splitting the sum
over $\omega_1$ in Eq.(32) into positive and negative parts, representing
$\omega_1$ as a time derivative which then is subjected to a partial 
integration, and finally performing the sums over $\omega_1$ and $n$
we obtain for the leading term of the nonlocal part $V_{NL}(\tilde{\phi})$ of $S_{ph}$
the following expression
\begin{equation}
V_{NL}(\tilde{\phi}) = {g^2\over 2} T^2 \int_0^\beta d \tau d \tau'
{{sin(2 \pi T(\tau- \tau'))} \over {1-cos(2 \pi T(\tau - \tau'))}}
\tilde{\phi}(\tau) {\partial{\tilde{\phi}(\tau')}\over{\partial \tau'}}
 G[\tilde{\phi}(\tau),\tilde{\phi}(\tau')],
\end{equation}
with
\begin{equation}
G[\tilde{\phi}(\tau),\tilde{\phi}(\tau')] = F[\tilde{\phi}(\tau)] + 
F[\tilde{\phi}(\tau')] -
{{\partial F} \over { \partial \tilde{\phi}(\tau')}}\cdot \tilde{\phi}(\tau').
\end{equation}
$V_{NL}$ vanishes for a constant field $\tilde{\phi}$ so it represents a nonlocal
correction to the local potential $V$. $V_{NL}$, on the other hand,
still depends on two arguments $\tau$ and $\tau'$ which would not occur
if a usual gradient expansion would be applicable. This feature is
caused by the nonanalyticity of $\Pi$ in Eq.(26): $|\omega_n|$ cannot be
represented just by one derivative but only by infinite many of them
causing the appearance of two arguments $\tau$ and $\tau'$ in Eq.(34).
We also would like to point out that the expression Eq.(34) is very 
similar to that derived by Hamann \cite{Hamann} for the Hubbard model using asymptotic
approximations for $G_0$ and avoiding a gradient expansion.

\section{The local electronic  Green's function $G_{loc}$}

The exact local Green's function $G_{loc}$ of the impurity problem
can be obtained from the relation
\begin{equation}
G_{loc}(\tau-\tau') = {{\delta log Z} \over {\delta G_0^{-1}(\tau-
\tau')}}.
\end{equation}
Carrying out the functional derivative we obtain
\begin{equation}
G_{loc}(\tau-\tau') = \int D \tilde{\phi} (G_0^{-1}(\tau_1-\tau_2)
-g\tilde{\phi}(\tau_1) \delta(\tau_1-\tau_2))^{-1}_{\tau-\tau'}
e^{S_{ph}} /Z.
\end{equation}
$S_{ph}$ is given by Eq.(12) where the $Tr\;log$ term has been
approximated by the local and nonlocal potentials $V$ and $V_{NL}$.
The simplest approximation for the functional integral in Eq.(37)
is obtained by considering $\tilde{\phi}$ as a classical field and
assuming the mass of the atoms to be infinite. The kinetic energy
of the atoms as well as the nonlocal part in $S_{ph}$ drops then out
and the functional integral becomes a usual integral. As a result we have
\begin{equation}
G_{loc}(\tau-\tau') = \int_\infty ^{\infty} d \tilde{\phi} (G_0^{-1}
-g\tilde{\phi})^{-1}_{\tau - \tau'} e^{-\beta V(\tilde{\phi})} /Z.
\end{equation}
Eq.(38) says that $G_{loc}$ is to be calculated for a general displacement
$\tilde{\phi}$ and then the result is averaged over all displacements
using a Boltzmann factor. In this approximation the electrons no longer
form a Fermi liquid and there is no quasiparticle peak in the electron
Green's function. For a finite mass of the atoms such an approach is 
valid if $T\gg\omega_0$. The dynamical mean field approximation on a Bethe
lattice amounts then to solve Eqs. (9), (20) and (38) self-consistently 
\cite{Millis1,Millis2}.      
\par
A general approximation scheme for path integrals valid especially at 
low temperatures
is based on expansions around stationary paths of the action $S_{eff}$.
These paths are solutions of the equation
\begin{equation}
{{\delta S_{eff}[\tilde{\phi}]} \over {\delta \tilde{\phi}(\tau)}} = 0.
\end{equation}
We will see that Eq.(39) has in general 3 solutions, two of them
being constant in time and one depending  on time. 
The surroundings of these solutions will yield the major contributions
to $G_{loc}$.

\subsection{Contributions to $G_{loc}$ from small fluctuations}

Time-independent solutions of Eq.(39) satisfy also
the equation
\begin{equation}
{{\delta V[\tilde{\phi}]}\over {\delta \tilde{\phi}}} = 0.
\end{equation}
Taking stability considerations into account the relevant solutions
of Eq.(40) correspond to the minima of $V$ which will be denoted by
$\tilde{\phi}_\alpha$, where $\alpha$ counts different minima. We expand now the 
field $\tilde{\phi}$ around each
minimum $\tilde{\phi}_\alpha$:
\begin{equation}
\tilde {\phi} = \tilde{\phi}_\alpha + \phi_\alpha'.
\end{equation}   
Inserting Eq.(41) into $S_{ph}$ we write
\begin{equation}
S_{ph}(\tilde{\phi}_\alpha + \phi'_\alpha) = S^{(0)}_{{ph},\alpha}
+S'_{{ph},\alpha},
\end{equation}
where $S^{(0)}_{{ph},\alpha}$ contains the kinetic energy of the atoms
and the quadratic terms in $\phi'_\alpha$ of $V$. $S'_{ph,\alpha}$
is due to the anharmonic part of $V$ around the minimum $\alpha$
and the nonlocal part $V_{NL}$. Expanding the inverse in Eq.(37)
in powers of $\phi'_{\alpha}$ as well as the exponential in powers of
$S'_{ph,\alpha}$ one obtains path integrals over the displacement field
containing an arbitrary number of powers in $\phi'_{\alpha}$. Since the
action in these path integrals is $S^{(0)}_{ph,\alpha}$ Wick's theorem
can be applied. Contributions to $G_{loc}$ from the minimum $\alpha$,
called $G_{\alpha}$ in the following, are thus obtained in terms
of usual diagrams due to the electron-phonon and anharmonic interactions.
As usual it is sufficient to consider self-energy diagrams 
$\Sigma_\alpha$ to $G_{\alpha}$. $G_{\alpha}$ and $\Sigma_\alpha$
are related by
\begin{equation}
G^{-1}_{\alpha} = G^{-1}_{0,\alpha} -\Sigma_\alpha,
\end{equation}
where $G_{0,\alpha}$ is the shifted unperturbed Green's function
defined by
\begin{equation}
G^{-1}_{0,\alpha} = G^{-1}_0 -g \tilde{\phi}_\alpha.
\end{equation}

Fig. 1 shows the lowest-order skeleton diagrams for $\Sigma_\alpha$
which are of second (diagram (a)) and of fourth order (diagram (b))
in $g$. All uncrossed diagrams are taken into account by diagram (a),
whereas diagram (b) represents the lowest order vertex correction to (a).
In conventional treatments of the electron-phonon coupling
it is argued that diagram (b) is smaller by the factor $\lambda \omega_0/t$
compared to the diagram (a). Whether this is true also in our case
is not apriori clear because Migdal's theorem relies heavily on 
properties of momentum integrals 
and phase space arguments for which there is no analogue in our case. \par

In order to judge the convergence of the perturbation expansion around
$\tilde{\phi_\alpha}$ we have calculated diagrams (a) and (b). One obtains
\begin{equation}
\Sigma^{(a)}_\alpha(i \omega_n) = -{g^2 \over {2 \omega_0}}
 b(\omega_0)G_{\alpha}(i \omega_n
+\omega_0) + {g^2 \over{2 \omega_0}}
 b(-\omega_0) G_{\alpha}(i \omega_n - \omega_0)
+g^2 \int d\epsilon \rho_\alpha(\epsilon) f(\epsilon){ 1 \over
{(\epsilon-i \omega_n)^2-\omega_0^2}},
\end{equation}
where $\rho_\alpha(\epsilon)$ is the spectral function of $G_{\alpha}$
and $b$ is the Bose distribution function. 

$\Sigma^{(b)}_\alpha$ is given by   
\begin{eqnarray}
\Sigma_\alpha^{(b)}& =& \sum_{s,t=\pm1} (-1)^{(s+t+2)/2}\;{g^4 \over {4 \omega_0^2}}
 \int d \epsilon
d \epsilon' d \epsilon'' {{\rho_\alpha(\epsilon) \rho_\alpha(\epsilon')
\rho_\alpha(\epsilon'')} \over {(-\epsilon+i \omega_n +s \omega_0)
(-\epsilon'' + i \omega_n +t \omega_0)}} \nonumber \\
&&\Bigl( {{A(\epsilon',s,t)}\over {i \omega_n +s \omega_0 +t \omega_0 - 
\epsilon'}} + {{B(\epsilon, \epsilon',t)} \over {\epsilon- \epsilon'
+t \omega_0}} + {{C(\epsilon',\epsilon'',s)} \over{\epsilon''+s \omega_0
-\epsilon'}} + {{ D(\epsilon, \epsilon', \epsilon'')} \over {\epsilon
-i \omega_n + \epsilon'' - \epsilon'}} \Bigr),
\end{eqnarray}
with 
\begin{equation}
A(\epsilon',s,t) = (b(t \omega_0) +f(\epsilon'))(b(s \omega_0) +f(
\epsilon' -t \omega_0)),
\end{equation}
\begin{equation}
B(\epsilon, \epsilon',t) = (b(t \omega_0) + f(\epsilon'))(f(\epsilon)
-f(-t \omega_0 + \epsilon')),
\end{equation}
\begin{equation}
C(\epsilon', \epsilon'',s) = (f(\epsilon'')-f(\epsilon'))(b(s \omega_0)
-b(\epsilon'-\epsilon'')),
\end{equation}
\begin{equation}
D(\epsilon,\epsilon',\epsilon'') = (f(\epsilon'')-f(\epsilon'))(
f(\epsilon) +b(\epsilon'-\epsilon'')).
\end{equation}
Figs. 2 and 3 present numerical results for the imaginary parts of diagrams
(a) (solid lines) and (b) (dashed lines) as a function of frequency. 
In Fig. 2 (3) we used the parameters $\lambda = 1.0 \; (1.375)$,
$\omega_0 = 0.1$,
and $T=0.01$, where all energies are expressed in units of $t$.
$G_{\alpha}$ has been calculated using the self-consistently determined
function $G_0$ described in the next section. The potential $V$
develops a double well around $\lambda = 1.2$ so that Figs. 2 and 3
correspond to the cases with one and two minima , respectively.
If Migdal's theorem would
hold diagram (b) should be smaller by one order of magnitude compared
to diagram (a). The Figures show that this is approximately the case
which validates the employed perturbation expansion. \par

The above result is nontrivial, especially, at large couplings
when the double well has been formed. As illustrated in the  
next section the main weight of $G_0$ resides in this case in a narrow,
central component at zero frequency. The characteristic energy of
$G_0$ thus may be low and comparable with $\omega_0$ making a simple
perturbation theory of $\Sigma_\alpha$ in terms of $G_{0,\alpha}$ impossible.
On the other hand, $G_\alpha$ consists at large couplings of
a broad peak of width $\sim t$ around $\tilde{\phi_\alpha}$.
The effective expansion parameter for an expansion of $\Sigma$
in terms of $G_\alpha$ can easily be determined
for large couplings where one can assume that $Im G_\alpha (\omega)$
is only non-zero for $\omega < -\omega_0$. Eqs.(45) and (46) reduce
then at $T=0$ to
\begin{equation}
\Sigma_\alpha^{(a)} \rightarrow {g^2 \over {2 \omega_0}}
G_\alpha(\omega+i \eta +\omega_0),
\end{equation}
\begin{equation}
\Sigma^{(b)}_\alpha \rightarrow {g^4 \over {4 \omega_0^2}}
 G_\alpha^2(\omega+i \eta +\omega_0)
G_\alpha(\omega +i \eta +2 \omega_0).
\end{equation}
The ratio $\Sigma_\alpha ^{(b)} / \Sigma_\alpha^{(a)}$ is thus
$\sim g^2/{\omega_0 t^2} = \lambda \omega_0/t$ which is in agreement with the
above numerical calculations. Nonvanishing potential
minima at $\tilde \phi_\alpha$ signalize the appearance of nonadiabatic
effects: at large couplings electrons occupy predominantly ionic
configurations with zero or 2 electrons at the atom. In each case the
lattices relaxes to a non-vanishing value similar as in the anti-adiabatic
case. As shown above, the perturbation expansions around the potential minima 
are, however, still controlled by the Migdal parameter $\lambda \omega_0 /t$.
\par

According to our previous discussion the nonlocal potential $V_{NL}$ 
is smaller by the Migdal parameter $\gamma$ compared to to the local
potential. To make this more quantitative we have calculated the
imaginary-part of the self-energy of $V_{NL}$ for a harmonic phonon
with frequency $\omega_0$ for a fixed, but general field value 
$\tilde{\phi}$. The result is shown in Fig. 4 for the two coupling strengths
used in Figs. 2 and 3 using again the self-consistent function $G_0$.
The position of the potential minima are indicated by arrows.
One concludes from Fig. 4 that the damping of the oscillations
around the potential minima is small enough in each case so that one may
neglect it. Finally, we have neglected anharmonic effects. This is an
excellent approximation at small and large couplings $\lambda$ but
not for intermediate couplings where the double well forms. A simple
prescription to take anharmonic effects into account which is based
on the exact atomic eigenfunctions in the potential $V$ will be
described in the next section. \par

\subsection{Instanton contribution to $G_{loc}$}

At half-filling the potential $V$ is symmetric with respect to $\tilde{\phi}
=0$. If $V$ has two minima they are equivalent and the potential barrier
between them is finite. The appearance of tunnelling processes
between the minima manifests itself in a third, time-dependent solution
of Eq.(39) \cite{Polyakov}. This instanton solution passes from one minimum to the
other as function of $\tau$ and is relevant at low temperatures where
this passage occurs within the interval $[0,\beta]$. One instanton
solution creates, however, a whole family of approximate solutions of
Eq.(39): The passage from one to the other minimum can take place
everywhere in the interval $[0,\beta]$; furtheremore, multi-instanton
configuration are also approximate solutions of Eq (39) and must be
taken into account. For each instanton
configuration we expand the inverse in Eq.(39) in powers of $\tilde{\phi}$
and sum then over all configurations using the rules derived 
in Ref.\cite{Polyakov}.
The first-order term in $\tilde{\phi}$ vanishes by symmetry. The
second-order self-energy can be represented by the diagram a) in Fig. 1
where ther phonon propagator is to be replaced by the instanton
propagator
\begin{equation}
D_I(i \omega_n) = \tilde{\phi}_1^2 {{1-e^{-\Delta E\beta}}\over
{1+e^{-\Delta E\beta}}} \cdot {{2 \Delta E} \over {\omega^2_n 
+(\Delta E)^2}}.
\end{equation}
$\tilde{\phi}_1^2$ is the squared field value of one of the two minima.
If $V_{NL}$ can be neglected $\Delta E$ denotes the tunnelling frequency
of one atom through the potential $V$ between the minima. In general,
$\Delta E$ contains an additional exponential factor with the
action due to $V_{NL}$ along one instanton in the exponent. $D_I$
has the form of a phonon propagator with the phonon frequency 
replaced by $\Delta E$ and an additional prefactor $\sim \Delta E / T$
at temperatures large compared to the tunnelling frequency. \par
The second-order instanton contribution to the self-energy becomes
\begin{eqnarray}
\Sigma_I(i \omega_n) &=& g^2 \tilde{\phi}_1^2{{1-e^{-\Delta E\beta}}\over
{1+e^{-\Delta E\beta}}}\Bigl( -b(\Delta E) G_0(i \omega_n + \Delta E)
+b(-\Delta E) G_0(i \omega_n - \Delta E) \nonumber \\
&&+ {g^2\over \omega_0} \int d \epsilon \rho(\epsilon)
f(\epsilon) {{\Delta E} \over {(\epsilon -i \omega_n)^2 -(\Delta E)^2}}
\Big),
\end{eqnarray}
where $\rho(\epsilon)$ is the spectral function of $G_0(\epsilon)$. 
In the limit $\Delta E \rightarrow 0$ $\Sigma_I$ approaches
$\tilde{\phi}^2_1 G_0(i \omega_0)$. Inserting this into Eq.(37)
one finds the contribution
\begin{equation}
G'_{loc} = {1 \over 2} ({1 \over{G_0^{-1}+g \tilde{\phi}_1}} +
{1 \over{G_0^{-1}-g \tilde{\phi}_1}}),
\end{equation}
to $G_{loc}$. 
$G'_{loc}$ is due to the integration over the paths along the 
two minima of $V$ which are also included in $\Sigma_I$. This
contribution has already been taken into account when integrating
over paths near the first two stationary path at $\tilde{\phi}_\alpha$
and thus should be subtracted again. The pure instanton contribution
to $G_{loc}$ is thus given in lowest-order perturbation theory by
\begin{equation}
G_{I} = {1 \over {G_0^{-1}-\Sigma_I}} - G'_{loc}.
\end{equation}
After the analytic continuation the imaginary part of $\Sigma_I$
deviates from $\tilde{\phi}_1^2 G_0(\omega+i \eta)$ only at small
frequencies $\omega \sim \Delta E$ where it rapidly decreases and 
approaches zero at zero temperature assuming that the characteristic energy
scale $t^\ast$ of $G_0$ is larger than $\Delta E$. Thus the
instanton contribution to $G_{loc}$ provides the finite value of
$G_{loc}$ of Fermi liquid theory at $\mu = 0$ in form of a Kondo resonance
peak with an exponentially small width determined by $\Delta E$. \par

The expansion of the denominator on the right-hand side of Eq.(37)
in powers of $g\tilde \phi$ seems to be problematic for
instanton field configurations because in this case $g\tilde \phi$ can
in general not be considered as small compared to $G_0^{-1}$.
This is also true if the perturbation series is resummed by means
of the self-energy $\Sigma_I$ and if the latter is approximated by its
lowest-order expression in $g$. In the remainder of this subsection
we will show that $G_{I}$ can be obtained in closed form
without using an expansion of the denominator in Eq.(37). The only
assumption is that the instanton gas is dilute, i.e., that self-energy
effects for the tunnel modes can be neglected. \par

Let us first disregard the nonlocal part $V_{NL}$ in the atomic
potential. The Hamiltonian for the impurity atom is then hermitean and
can be diagonalized  yielding two low-lying states which are split
by tunnelling processes. Disregarding small oscillations around the
extremal paths the evaluation of the path integral by instanton solutions
corresponds to the solution of the Schroedinger equation for the
two lowest-lying states of the Hamiltonian. Thus the instanton part
of the electron Green's function, $G_I$, can also be obtained 
from an effective two-level model with the Hamiltonian
\begin{equation}
H_I = {{\Delta E}\over 2} (X^{11} - X^{00}).
\end{equation}
$X^{11}$ and $X^{00}$ are Hubbard operators defined as the projection
operators $|1><1|$ and $|0><0|$, respectively, where $|0>,|1>$ are
the two lowest states of the atomic Hamiltonian and $\Delta E$ their
splitting. The field variable $\tilde{\phi}$ becomes
\begin{equation}
\tilde{\phi} = \tilde{\phi}_0(X^{01}+X^{10}),
\end{equation}
where $\tilde{\phi}_0$ is the matrix element $<0|\tilde{\phi}|1>$
 which can be identified
with the previous $\tilde{\phi}_{1}$ in a good approximation. The 
nondiagonal Hubbard operators $X^{01}$ and $X^{10}$ are given by 
$|0><1|$ and $|1><0|$, respectively. The
(imaginary) time evolution of the nondiagonal operators due to $H_I$ is
\begin{equation}
X^{01}(\tau) = e^{-\Delta E \tau} X^{01}(0),
\end{equation}
\begin{equation}
X^{10}(\tau) = e^{\Delta E \tau} X^{10}(0).
\end{equation} 
Using Eqs.(59) and (60) the displacement Green's function $D_I(i \omega_n)$
can be calculated in a straightforward way and yields exactly the
previous expression Eq.(53). Taking also $V_{NL}$ into account
changes the extremal instanton configurations. However, in the dilute 
instanton approximation, only the action along the extremal path enters.
This means that $\Delta E$ acquires a different meaning in the effective
two-level model but otherwise there is no change. \par

In the following it is convenient to use a description based on a
Hamiltonian and not on a Lagrangian. To this end we consider the
following effective Hamiltonian
\begin{eqnarray}
H_{eff}&=& \sum_{l\sigma} \tilde{\epsilon}_l a^\dagger_{l \sigma} a_{l \sigma}
+ \sum_{l \sigma} V_l(a^\dagger_{l \sigma} c_\sigma +c^\dagger_\sigma 
a_{l \sigma})
-\mu \sum_{\sigma} c^\dagger_\sigma c_\sigma \nonumber \\
&+& {{\Delta E}\over 2} (X^{11}-X^{00}) +g\phi_0(X^{01}+X^{10})\sum_\sigma
c^\dagger_\sigma c_\sigma.
\end{eqnarray}
The first three terms in $H_{eff}$ represent an Anderson Hamiltonian
for impurity electrons, described by $c^\dagger_\sigma,c_\sigma$,
interacting with bath electrons, described by $a^\dagger_{l \sigma},
a_{l \sigma}$. As shown in Ref.\cite{Kotliar} elimination of the bath electrons
allows to represent a general function $G_0$ in the
effective impurity action. The fourth term represents $H_I$, and
the last term the interaction between impurity electrons and the tunnel
modes. \par
Neglecting self-energy effects in the tunnel modes the equation of motion 
for the composite operator $X^{pq}(\tau) c_\sigma(\tau)$ is,
using $H_{eff}$,
\begin{equation}
{{\partial} \over {\partial \tau}}(X^{pq} c_\sigma ) = \left\{
\begin{array}{l r}
X^{00}{{\partial c_\sigma}\over {\partial \tau}}     &X^{01}(-\Delta E 
c_\sigma+{{\partial c_\sigma} \over {\partial \tau}}) \\
X^{10}(\Delta E c_\sigma+{{\partial c_\sigma}\over{\partial \tau}})
&X^{11} {{\partial c_\sigma} \over {\partial \tau}}
\end{array}
\right\}.
\end{equation}
with $p,q=0,1$. The equations of motion for the operators $c_\sigma$
and $a_{l\sigma}$ are
\begin{equation}
{{\partial c_\sigma} \over {\partial \tau}} = \mu c_\sigma -
g\phi_0(X^{01}+X^{10})c_\sigma -\sum_{l} V_l a_{l\sigma},
\end{equation}
\begin{equation}
{{\partial a_{l \sigma}} \over {\partial \tau}} = - \tilde{\epsilon}
a_{l \sigma} -V_l c_\sigma .
\end{equation}
Inserting Eq.(64) into Eq.(63) yields
\begin{equation}
\int_0^\beta d{\tau'} \Bigl[ \delta(\tau - \tau')(-{\partial \over 
{\partial \tau'}} +\mu) -\Lambda(\tau - \tau')\Bigr] c_\sigma(\tau') =
-g \phi_0 (X^{01}+X^{10}),
\end{equation}
with the function
\begin{equation}
\Lambda(\tau - \tau') = T \sum_{l,n} V_l^2 {{e^{-i \omega_n(\tau-\tau')}}
\over{i \omega_n -\tilde{\epsilon}}}.
\end{equation}
Inserting Eq.(65) into Eq.(62) yields
\begin{equation}
\int_0^\beta d {\tau'} \Bigl[ \delta(\tau- \tau') ({\partial \over
{\partial \tau'}} - \mu) +\Lambda^{pq}(\tau - \tau') \Bigr] X^{pq}(\tau')
c_\sigma(\tau') = \sum_{p'q'} M_{pq,p'q'} X^{p'q'}(\tau) c_\sigma(\tau).
\end{equation}
The only non-vanishing elements of $M$ are
\begin{equation}
M_{00,01}=M_{01,00}=M_{10,11}=M_{11,10}= -g\phi_0,
\end{equation}
\begin{equation}
M_{10,10}=-M_{01,01}=\Delta E.
\end{equation}
We also used the definitions
\begin{equation}
\Lambda^{00}(\tau-\tau') = \Lambda^{11}(\tau-\tau')=\Lambda(\tau-\tau'),
\end{equation}
\begin{equation}
\Lambda^{01}(\tau-\tau') = \Lambda(\tau-\tau')e^{-\Delta E (\tau-\tau')},
\end{equation}
\begin{equation}
\Lambda^{10}(\tau-\tau') = \Lambda(\tau-\tau')e^{\Delta E (\tau-\tau')}.
\end{equation}
\par
Let us now define the following Matsubara Green's functions
\begin{equation}
G({{pq} \atop {\tau-\tau'}}) = -<T(X^{pq}(\tau)c_\sigma(\tau)
c^\dagger_\sigma(\tau'))>.
\end{equation}
They satisfy the equation of motion
\begin{equation}
\int_0^\beta d{\tau''} G_0^{-1}({{pq}\atop{\tau-\tau''}}) G({{pq}\atop{
\tau''-\tau'}}) + \sum_{p'q'}M_{pq,p'q'} G({{p'q'}\atop{\tau-\tau'}})
=\delta(\tau-\tau')<X^{pq}>,
\end{equation}
with the definition
\begin{equation}
G_0^{-1}({{pq}\atop{\tau-\tau'}}) = \delta(\tau-\tau')(-{{\partial}
\over{\partial \tau''}} +\mu) - \Lambda^{pq}(\tau-\tau').
\end{equation}
Comparing Eqs.(70) and (75) with Eqs.(14) and (15) of Ref.\cite{Kotliar}
one finds that
\begin{equation}
G_0({{00}\atop{\tau-\tau'}})= G_0({{11}\atop{\tau-\tau'}})
= G_0(\tau-\tau').
\end{equation}
Furthermore, one can easily verify that Eq.(75) yields
\begin{equation}
G_0({{10}\atop{\tau-\tau'}})=G_0(\tau-\tau')e^{\Delta E (\tau-\tau')},
\end{equation}
\begin{equation}
G_0({{01}\atop{\tau-\tau'}})=G_0(\tau-\tau')e^{-\Delta E (\tau-\tau')}.
\end{equation}
Using these results and eliminating the non-diagonal Green's functions
one finds that Eq.(74) is equivalent with the two equations
\begin{eqnarray}
G({{00}\atop{\tau-\tau'}}) = G_0(\tau-\tau')<X^{00}> \nonumber \\
+g^2\phi_0^2 
\int_0^\beta d\tau'' d\tau''' G_0(\tau-\tau'') G_0(\tau''-\tau''')
e^{-\Delta E (\tau''-\tau''')}G({{00}\atop{\tau'''-\tau'}}),
\end{eqnarray}
\begin{eqnarray}
G({{11}\atop{\tau-\tau'}}) = G_0(\tau-\tau')<X^{11}> \nonumber \\
+g^2\phi_0^2 
\int_0^\beta d\tau'' d\tau''' G_0(\tau-\tau'') G_0(\tau''-\tau''')
e^{\Delta E (\tau''-\tau''')}G({{00}\atop{\tau'''-\tau'}}).
\end{eqnarray}
\par
Eqs.(79) and (80) can easily be solved in $\omega$-space. One obtains
then an explicit expression for the wanted Green's function
\begin{equation}
G(i\omega_n) = G({{00}\atop{i\omega_n}})+G({{11}\atop{i\omega_n}}).
\end{equation}
From $G$ one obtains the following expression for the self-energy
$\Sigma_I$:
\begin{equation}
\Sigma_I(i \omega_n) = {{<X^{00}>\Pi^{(1)}(i\omega_n)+<X^{11}>\Pi^{(2)}
(i\omega_n)-G_0(i\omega_n)\Pi^{(1)}(i\omega_n)\Pi^{(2)}(i\omega_n)}\over
{1-G_0(i\omega_n)(<X^{00}>\Pi^{(2)}(i\omega_n)+<X^{(11)}>\Pi^{(1)}
(i\omega_n))}},
\end{equation}
with
\begin{equation}
\Pi^{(1,2)}(i\omega_n) = g^2\phi^2_0T\sum_mG_0(i\omega_m)
{{e^{\mp\Delta E \beta}-1} \over{ i\omega_n - i\omega_m \mp \Delta E}},
\end{equation}
where the upper (lower) signs in front of $\Delta E$ refer to the superscripts
1 (2). In the weak-coupling limit $g\ll1$, but also in the high-temperature
limit $\Delta E \beta \ll1$, Eqs.(82) and (83) reduce exactly to 
the previous result Eq.(54). The same also holds at zero temperature
where $<X^{00}>=1$ and $<X^{11}>=0$. Eqs.(54) and (82) have, for a
finite $\Delta E$, also the same low temperature limit which is
a rather astonishing result.
\par

Taking all together the local Green's function has been obtained
in general as a sum of three contributions
\begin{equation}
G_{loc} = {Z_1 \over Z} G_{1}+ {Z_2 \over Z} G_{2} + G_{I}.
\end{equation}
$Z_\alpha$ is the contribution of the minimum $\alpha$ to the 
partition function $Z$. 
At small couplings $V$ has only one minimum at $\tilde{\phi}=0.$ There 
exists then only the first
term in Eq.(84) with $Z_1/Z=1$  and the perturbation theory can be
carried out in powers
of $\tilde{\phi}$ as usual. Beyond a critical coupling strength $V$
develops two minima. The perturbation theory must then be performed
around each minimum separately yielding the first two terms in Eq.(58).
There are in addition tunnelling processes between the two minima 
causing the third term in Eq.(84). 
The appearance of the prefactors $Z_\alpha/Z$ in Eq.(84) is due
to the fact that the use of the linked cluster theorem in the expansion
around $\alpha$ cancels the contribution $Z_\alpha$. The definition
Eq.(37) of $G_{loc}$ implies then the above prefactors. The use of
the linked cluster theorem in calculating $G_I$ means that there is no 
prefactor in front of $G_I$ in Eq.(84). In general $Z=Z_1+Z_2+Z_I$
where $Z_I$ is the instanton contribution to $G_{loc}$. A convenient
way to determine $Z_I$ is to apply the sum rule to Eq.(84). In practice
$Z_I$ is small compared to $Z$ so that $Z_1/Z =Z_2/Z = 1/2$ can
be used in a very good approximation.

\section{Numerical results}

In this section we discuss the results obtained by solving numerically
the set of dynamical mean-field equations on the Bethe lattice. 
$<\phi>$ was determined from Eq.(15) from the outset. Then we proceeded 
in an iterative way, starting with an educated guess
for $G_{0}$. The effective potential $V({\tilde \phi})$ was then calculated
from Eq.(21). The calculation of the local Green's function $G_{loc}$,
given by Eq.(84), amounted to calculating the self-energies $\Sigma_\alpha$
and $\Sigma_I$. To account also for anharmonic effects we determined
the stationary solutions of the Schr\"odinger equation for one atom
in the potential $V$:
\begin{equation}
\frac {\partial^{2}\psi_{r}(\tilde{ \phi})}{\partial\tilde{ \phi}^{2}}+
2[E_{r}-V(\tilde{ \phi} )]\psi_{r}(\tilde{ \phi})=0 .
\end{equation}
$r$ labels the different states with energy $E_r$ and eigenfunction
$\psi_r(\tilde \phi)$. In the weak-coupling case, where $V$ has only one
minimum, we obtained the phonon propagator $D$ using the exact
energies and eigenstates,
\begin{eqnarray}
D(i\omega_{n} )&=&\frac{1}{Z_{ph}}\sum_{r \neq s}e^{-\beta E_{s}}
\frac{ |\langle s|\tilde{ \phi} |r \rangle|^{2}
(1-e^{-\beta\Omega_{rs }})}{i\omega_{n}-\Omega_{rs}},
\end{eqnarray}
with $\Omega_{rs} = E_r - E_s$ and $Z_{ph} = \sum_r e^{-\beta E_r}$.
Only the first term in Eq.(84) is present and the corresponding self-energy
is calculated from the diagram in Fig.1a) using Eq.(86) for the
wavy line. As a result anharmonic effects were fully taken into
account whereas higher-order skeleton diagrams and also $V_{NL}$ can be 
neglected as shown in the previous section.
\par
In the strong-coupling case $V$ exhibits two minima. We treated
the two lowest levels as a tunnel system and omitted the transitions between
them in the phonon propagator Eq.(86). Furthermore, the remaining 
transitions 
in Eq.(86) can be ascribed either to the left or the right well because
at least one wave function involved in the transition is localized.
We found it most convenient to accomplish this splitting of $D$ into
left and right parts by dividing 
the matrix element $<s|\tilde \phi|r>$ into contributions with 
${\tilde \phi}>0$ and ${\tilde \phi}<0$ using also appropriate
normalization factors. The self-energies $\Sigma_\alpha, \alpha=1,2$, 
were then calculated using again the diagram in Fig.1 (a) with the split
phonon propagator as wavy line. Our procedure becomes exact at large
couplings and also provides a very smooth transition to the one-well
case. 
Finally $G_I$ was determined from Eq.(82). Since Eq.(82)
approaches the simpler expression Eq.(52) in several limiting cases we
found it sufficient to use Eq.(52) and also to neglect $V_{NL}$
in calculating $\Delta E$. Having obtained $G_{loc}$ we determined a new
$G_0$ from Eq.(9) and started a new cycle until convergency was reached.

\subsection {Atomic properties}

Fig. 5 shows the self-consistent potential $V$ (in unit of the hopping
t) as a function of the
dimensionless phonon field $\xi = {\tilde \phi} \omega_0/t^{1/2}$
for $T=0.1 \omega_0 = 0.01 t$. The dotted line corresponds to $\lambda = 0.5$,
i.e., the weak-coupling case where $V$ exhibits one minimum. The dashed curve
refers to $\lambda = 1.2$ and corresponds to the case where the potential
is very flat around $\xi \sim 0$ and the double well structure has
just been formed. The double well becomes more and more pronounced with
increasing $\lambda$ as shown by the dot-dashed and solid lines. The
dotted line in Fig. 6 shows the position of the
potential minimum $R=\tilde{\phi}_{1}\, \omega_{0}/t^{1/2}$ as a function of $\lambda$ for $T=0.1 \omega_0 = 
0.01 t$. The circles are calculated values which have been joined smoothly
by the dotted curve. The Figure illustrates how rapidly $R$ increases
above the threshold value $\lambda_c \sim 1.15$. At large $\lambda$'s
$R$ approaches the curve $\sqrt{\lambda}$ shown in Fig. 6 as a solid line.
The squares in Fig. 7, joined smoothly by a dashed line, represent the
difference $(V_{max}-V_{min})/10$ in units of $t$ as a function of $\lambda$.
$V_{max}$ and $V_{min}$ are the values $V(0)$ and the minimum value of $V$,
respectively. $V_{max}-V_{min}$ approaches at large $\lambda$  
the curve $\lambda /2$. The filled circles in Fig. 7 describe the
energy difference $\Omega_{10}$ between the two lowest eigenvalues,
i.e., a phonon frequency. This frequency is at $\lambda=0$
equal to the bare phonon frequency $\omega_0 = 0.1 t$. It softens
with increasing $\lambda$ and tends to a very small value when the double 
well has been formed. This ``soft'' phonon does not produce a structural
phase transition for $\lambda > \lambda_c$ but switches over smoothly to the 
tunnel mode connecting the two potential minima. The corresponding frequency
is practical zero on the energy scale of Fig. 7. The empty circles in Fig. 7
describe the excitation energy $\Omega_{20}$. At $\lambda = 0$ it is equal
to $2 \omega_0$, softens appreciably with increasing $\lambda$ up to
the critical value $\lambda_c$, and then recovers approaching $\omega_0$
from below at large $\lambda$'s. Fig. 8 depicts three curves for the
probability function $P(\xi)$ to find an atom at a distance $\xi$ as
a function of $\xi$. $P$ shows for $\lambda = 0.5$ a one-peak structure
which broadens and finally splits into two peaks with increasing $\lambda$.
Finally, Fig. 9 shows the electron-phonon correlation function $C_0 =
-g/t<{\tilde \phi}n>$ versus $\lambda$ for $T=0.1 \omega_0 =0.01 t$. In the
weak-coupling limit $C_0$ is very small which means that the atomic 
displacements
are rather independent from electronic fluctuations and depend only on the 
average electronic density. This changes dramatically for $\lambda > 
\lambda_c$ where the atomic displacement depend strongly on the
momentary electronic occupation of the atomic site. \par

The interpretation of the above results is rather straightforward.
In the weak-coupling case, for instance at $\lambda = 0.5$, the atoms
are according to their probability distribution function located mainly
near $\xi \sim 0$ and have only a small mean square displacement 
from this position. From Eq.(9) follows that the charge fluctuations
around the average value $n=1$ are then also small. This is the
situation corresponding to the description of a metal in the Migdal
approximation: The electrons hop very often between sites during one
atomic oscillation. As a result the atoms oscillate around their
unperturbed equilibrium positions with only slightly changed frequencies.
With increasing $\lambda$ the potential energy of the electrons becomes more 
and more important compared to their kinetic energies and the electrons
prefer to stay in doubly occupied sites. Since the average occupation
of one site is forced to be one this means that each site is occupied 
for long times
either with two or with zero electrons. The atoms adapt to each of this ionic
configuration by relaxing either to the left or right potential minimum.
If one assumes that the atoms have enough time to relax completely to the
static equilibrium for each ionic configuration $R$ is given by 
$\sqrt{\lambda}$. The solid line in Fig. 6 shows indeed that $R$
approaches $\sqrt{\lambda}$ at large $\lambda$'s. Using a similar 
consideration 
$V_{max}-V_{min}$ should approach the curve $\lambda /2$, which 
is in agreement with Fig. 7. Finally, one expects that at large couplings the
atoms oscillate more or less with the unperturbed frequency around the new
equilibrium positions for each ionic configuration. This conclusion is also 
supported by the curve for $\Omega_{20}$ in Fig. 7.
\par
Our path integral method coincides with the usual Migdal-Eliashberg
perturbation expansion in the weak-coupling limit. In the strong-coupling
case our method resembles in several respects treatments based on the
Lang-Firsov transformation \cite{Alex-Mott,Lang-Firsov}. 
One common feature is that the atoms
acquire new equilibrium positions depending on the momentary electronic
configuration. In the Lang-Firsov approach the new equilibrium values 
are given by $\xi = \pm \sqrt{\lambda}$. Fig. 6 shows for the Migdal
parameter $\omega_0/t=0.1$ that our expansion points $\pm R$ are near
to these values for $\lambda \gg \lambda_c$.
The perturbation expansion around these points, however, is ruled by
different expansion parameters: In the Lang-Firsov case \cite{Lang-Firsov} by $1/
\lambda$ and in our case by $\lambda \cdot \omega_0 /t$. We also 
note that the appearance of expansion points different from zero
signalizes the breakdown of the Migdal-Eliashberg expansion:
In case of a double-well the information obtained  by expansions around the
minima cannot retrieved by low-order diagrams of an expansion around
the local maximum at $\xi = 0$. Our calculations imply that the
validity of the Migdal-Eliashberg perturbation expansion is not
solely governed by the condition $\lambda \cdot \omega_0/t \ll1$,
as often is assumed, but also by $\lambda \leq \lambda_c$, where $\lambda_c$
lies around 1.2 for a small Migdal parameter $\omega_0/t$. Based
on the Lang-Firsov transformation the condition $\lambda \leq \lambda_c$
for the validity of the Migdal-Eliashberg expansion has also been stressed
in Ref.\cite{Alex-Mott}. 

\subsection {Electronic properties at low temperatures}
In this section we present numerical results
for the electronic properties of the Holstein model. In the last section
we have seen that the crossing from the weak to the strong-coupling regime 
is characterized by a drastic change in the effective atomic potential 
due to the appearance of anharmonic effects and the formation of a double 
well. These effects also cause typical features in the spectral
properties of electrons. Fig. 10 shows the evolution of the spectral
function $\rho(\omega) = -Im\;G_{loc}(\omega)/\pi$ with coupling
strength $\lambda$ as a function of frequency $\omega$ for a finite
but low temperature $T=0.1 \omega_0 = 0.01 t$. In case of the weak-coupling
curve with $\lambda = 0.5$ $\rho(\omega)$ is somewhat broadened
compared to the semicircular density of free electronic states.
In the metallic state $\rho(0)$ should always be unrenormalized and equal
to $1/\pi$ at $T=0$. Our value for $\rho(0)$ is slightly smaller
than $1/\pi$ because of the finite temperature. For the next lower curve
with $\lambda =1.25$ the double well has already been formed. Nevertheless,
$\rho(\omega)$ still consists of a broadened peak centered at $\omega = 0$.
Atop of this broad structure is a narrow component caused also by the
tunneling of atoms between the potential minima. With increasing $\lambda$
the broad peak for $\rho$ splits symmetrically into two peak centered
around $\pm \sqrt{\lambda} R$ with widths comparable to the width of the density of
bare states. For $\lambda < 1.5$ the system is still in the metallic state,
however, $\rho(0)$ is much smaller than the canonical value $1/\pi$.
The reason for this is that the tunneling contribution mainly
determines the value of $\rho(0)$. This contribution, however, 
is heavily quenched if the temperature is larger or comparable with
the tunnelling  energy $\Delta E$. The latter becomes exponentially
small at large $\lambda$'s. For our temperature $T=0.1 \omega_0$
the tunnel contribution to $\rho(\omega)$ is invisible small in plots
like Fig. 10 when the two peaks already separate at around $\lambda \sim 1.5$.
The interpretation of the curves in Fig. 10 is simple:
For $\lambda < \sim 1.3$ charge fluctuations at a site around the 
average value 1 are small. For values for $\lambda$ between
$\sim 1.3$ and $1.5$ the charge fluctuations become larger and
larger and the electronic state at a site consists of a coherent
superposition of a doubly and an unoccupied state. At $\lambda 
\sim 1.5$ the coherency is practically lost leading to an incoherent
arrangement of doubly and unoccupied sites. 
\par
Fig. 11 shows the negative imaginary part of $G_0$. For small $\lambda$'s
this function resembles the density of bare states. With increasing $\lambda$
it develops a strong central component which looks like a needle for
$\lambda > 1.5$. At the same time spectral weight is shifted to the
upper and lower ends of the spectrum which gradually transforms into
an upper and lower subband. With increasing $\lambda$ these subbands
move away from the center and lose spectral weight. Fig. 12 shows the
evolution of $-Im\;\Sigma(\omega)$ with $\lambda$ as a function of
$\omega$ for $T=0.1 \omega_0 = 0.01 t$. For the values of $\lambda$
on the left panel of this figure $-Im\;\Sigma$ is essentially proportional 
to the density of bare states except at small frequencies where it
drops to small values reflecting the Fermi liquid ground state.
The region of small values decreases with increasing $\lambda$
due to the softening of the phonon and the appearance of the tunnel mode.
For the larger values for $\lambda$ used in the right panel the spectral
weight accumulates in a central component and two sidebands which
move away from the center. 
\par
The low-frequency behavior of the real part of the self-energy
can be characterized by an effective mass $m^*$ defined by
 \begin{equation}
m^{*}/m=\left.1-\frac{d Re \Sigma(\omega)}{d\omega}\right|_{\omega=0} .
\end{equation}
Figure 13 shows $m^*/m$ as a function of $\lambda$
for $T=0.1 \omega_0 = 0.01 t$. At small $\lambda$ the mass enhancement
is given by $1+\lambda /\pi$. Above $\lambda \sim 0.8$ $m^*$ increases
much stronger than linearly which announces localization and the
transition to the insulating state. The metal-insulator transition 
can also be characterized by the
frequency dependence of the ac conductivity $\sigma(\Omega)$
given by \cite{Kotliar}
\begin{equation}
\sigma(\Omega)=\frac{2e^{2}t \pi}{\Omega}\int d\epsilon N(\epsilon)
\int d\omega A(\omega,\epsilon)A(\omega+\Omega,\epsilon)( f(\omega)-f(\omega+\Omega)) .
\end{equation}
$A(\omega,\epsilon)=-Im\;G(\omega,\epsilon)/ \pi$ is the one particle 
spectral function of the lattice Green's function.
$\sigma(\omega)$ is shown in Fig. 14 for four different values of $\lambda$.
The large Drude peak seen for $\lambda =1.2$ splits for $\lambda = 1.36$
into a small peak at $\omega = 0$ and a broader peak at a finite value
of $\omega$. Beyond $\lambda \sim 1.4$ the Drude peak has virtually
vanished and all spectral weight has been accumulated in the broad peak
caused by transitions between the two subbands.
\par
Fig. 15 shows the frequency dependence of the spectral function 
$A(\omega, \epsilon)$ of the 
lattice Green's function at the Fermi surface  $\epsilon=0$. The thick solid line
corresponding to $\lambda = 1.36$ exhibits the quasiparticle peak at $\omega
= 0$ and a rather well developed two-peak structure of the incoherent
background. The thin solid line represents the same spectral function
calculated within the Migdal-Eliashberg perturbation
expansion where the phonon field is expanded around ${\tilde \phi}
= 0$. The same parameter has been used as well as the exact eigenstates
in the potential $V$. The quasi-particle peak is in this case more pronounced
and the incoherent background is structureless. Our path
integral approach shows that the point ${\tilde \phi} = 0$ is an
unstable extremal point in this case and that the two potential
minima are stable extremal points suitable for expansions. This clearly
means that the thin line in Fig. 15 is incorrect and that the Migdal-
Eliashberg expansion has broken down in spite of the small Migdal
parameter $\gamma=0.1$ used in this figure. For $\lambda = 2$ both the
thick and thin dashed-dotted curves exhibit no longer visible quasiparticle
peaks at $\omega = 0$. The Migdal-Eliashberg expansion, however, is unable
to reproduce the two-peak structure of the incoherent background
but instead shows a broad and structureless peak. Though a
double well potential has been formed already for $\lambda = 1.25$
the two approaches give similar results in this case, mainly because 
the first excited state is above the barrier and the trajectories of the atom
are rather delocalized. This means that 
for a Migdal parameter of 0.1 the Migdal-Eliashberg expansion
breaks down for $\lambda \geq 1.25$.
\par

\subsection {Electronic properties at high temperatures}

To complete the overview of the electronic properties we show here also
results obtained in the limit $T \gg \omega_{0}$. This limit is equivalent
to the limit $\omega_0 \rightarrow 0$ and corresponds to the case where the
phonons behave like a quasi-static, classical field. $G_{loc}$ can then be 
calculated from Eq.(38) and an explicit high-temperature expansion of Eq.(45) 
and (46) shows that the corrections are of the order $\lambda \omega_0 /T$.
On the other hand, the validity of our general quantum-mechanical approach 
requires $\lambda \omega_0 /t \ll1$ and $\lambda T/t \ll 1$. The first
inequality follows from Eqs.(51) and (52), the second one by expanding
Eqs.(45) and (46) at high temperatures after taking the limit 
$\omega_0 \rightarrow 0$. \par
  The above expansion parameters suggest that the two approaches
should yield similar results if $\omega_0 \ll T \ll t$. To check this
we calculated $\rho(\omega)$ for $T=2.0 \omega_0 = 0.2 t$.
The results are shown in Figs. 16 and 17. For $\lambda = 0.5$
the above expansion parameters are small and the two corresponding
curves indeed are practically coincident. This agrees with the observation
that in the quantum-mechanical approach $Im\;\Sigma(\omega = 0)$ is 
already so large that the quasi-particle peak in the
spectral function of the lattice Green's function has completely vanished.
For $\lambda > 1.2$ similar double well potentials form in the two cases.
In the quantum-mechanical treatment the instanton contribution can safely
be neglected at these high temperatures and thermally activated processes
seem somewhat impeded by the level quantization compared to the classical
case. As a result the splitting of the band into two subbands occurs
more slowly in the classical, quasi-static case than in the quantum-mechanical
case. For $\lambda = 4.$ $\rho(\omega)$ shows in both cases well-splitted
bands with a pseudo-gap in between where the density is finite but
exponentially small. In Fig. 16 the shape of the two peaks resembles
the bare density of the Bethe lattice whereas in Fig. 17 the shape is
more Gaussian with well-developed tails into the pseudo-gap.
The two Figures illustrate that the quantum-mechanical treatment
approaches smoothly the classical case with infinite heavy atoms
at high temperatures and agrees with the latter also quantitatively if
all expansion parameters are small.

\section{Conclusions}

We have studied the Holstein model in infinite dimensions at half-filling 
in the normal state. In contrast to recent semiclassical treatments
we consider the atomic displacements as a quantized, dynamical field
which enables us to recover Fermi liquid properties at low temperatures
in the metallic state. Our approximation scheme is based on expansions
of path integrals around extremal paths of the action. 
This method allows to treat arbitrary strengths of the coupling
constant $\lambda$ on the same footing. For each
stationary path a Migdal-Eliashberg-like expansion is generated which
converges well both for weak and strong couplings. Beyond a critical value of 
$\lambda$ there is also a time-dependent, extremal instanton path 
associated with
the tunneling of atoms between equivalent minima in the effective atomic
potential. Assuming that the instantons are dilute a perturbation
expansion of the electronic self-energy in terms of unrenormalized instanton 
propagators has been carried out and shown that for most purposes only the
lowest-order, bare term has to be included.

We found that the usual Migdal-Eliashberg perturbation expansion breaks down 
around $\lambda \sim  1.25$. This breakdown is caused by the
appearance of more than one extremal path of the action. Related to this
is that the original expansion point at $\tilde {\phi} = 0$ for the
phonon field $\tilde {\phi}$ corresponds no longer to a minimum but to
a local maximum of the effective potential. The following picture
emerges from the calculations: For $\lambda < 1.2$ the system is in a
metallic state which can be described by Migdal-Eliashberg theory.
Phonon renormalizations are then weak and the electronic mass enhancement
is linear in $\lambda$. For $\lambda >1.2$ the adiabatic
approximation breaks down. The system stays mainly in 
ionic states with either two or zero electrons. Each state has its own
displacement pattern and instantaneous equilibrium position. The system is
still metallic due to atomic tunneling processes between the potential
minima, however, the electronic spectral functions develop a two-peak
structure which cannot be obtained in the Migdal-Eliashberg theory. Increasing
$\lambda$ further the electronic band splits into two subbands,
the system becomes effectively insulating and behaves like a mixture of
unoccupied and doubly occupied states. The semiclassical treatment and
also the exact solution for vanishing hopping show that this ``insulating''
state is for any finite temperature not truly insulating: The density of
electronic states between the two subbands is exponentially small
but not exactly zero.

{\bf Acknowledgement:} Both authors thank S. Ciuchi for stimulating and 
useful discussions. 
The first author acknowledges useful discussions with E. Cappelluti.

\begin{figure}
\vspace{+5cm}
\epsfysize=8.8in
\hspace*{1cm}
\epsffile{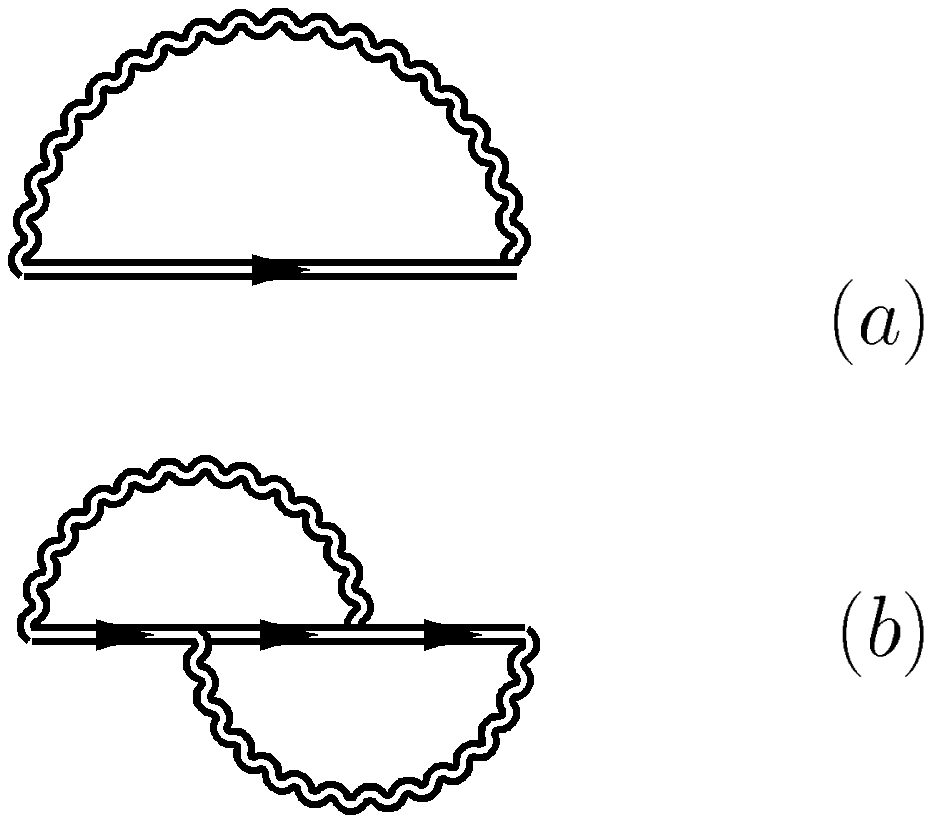}
\vspace{-10cm}
\caption
{Skeleton graphs for the electronic self-energy of second (a) and fourth
order (b) in g. 
The double solid lines denote the dressed electronic Green's
function and the double wavy lines denote the full phonon propagator.}
\label{Feyn}
\end{figure}
\newpage

\begin{figure}
\centerline{\psfig{figure=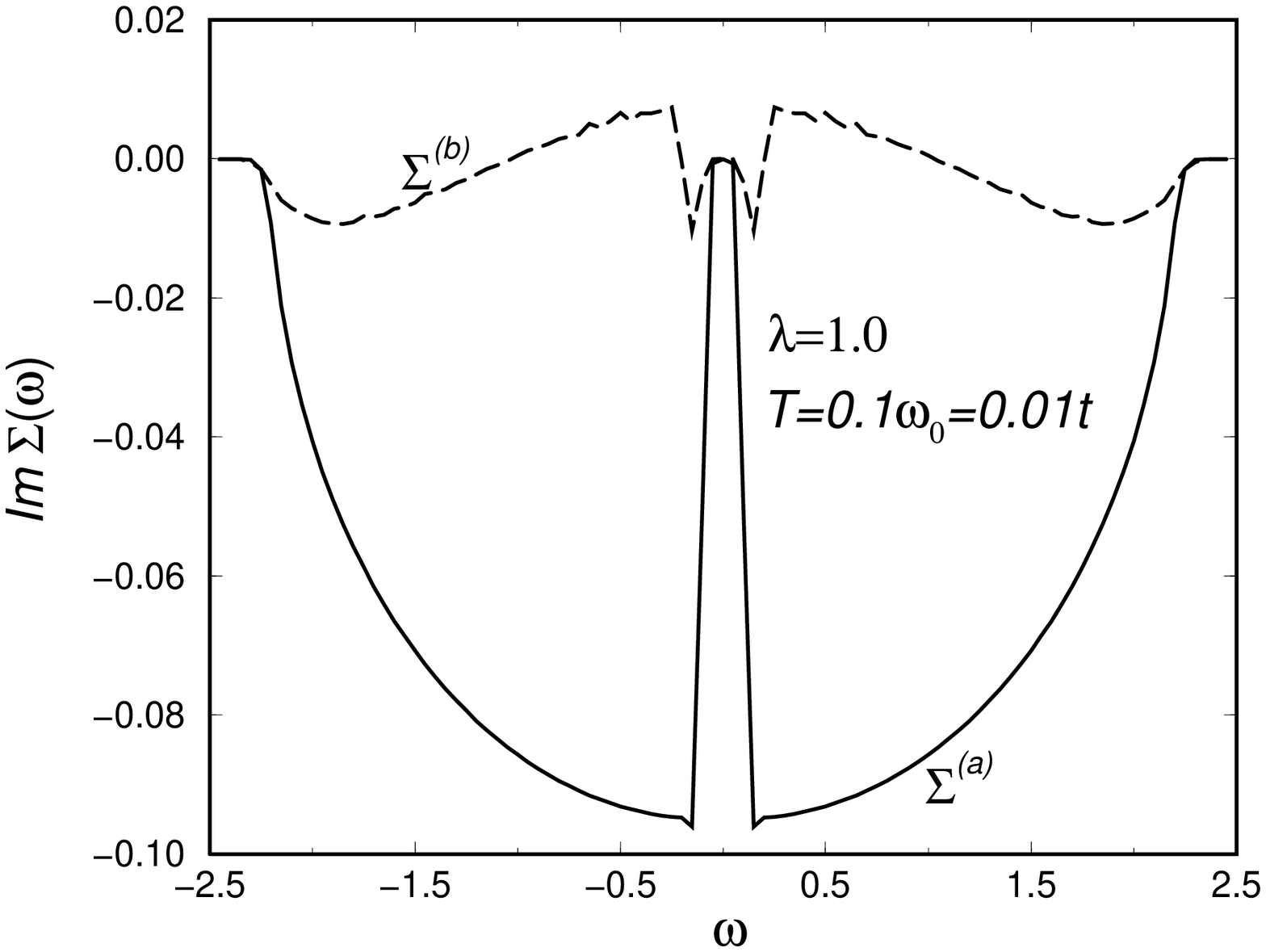,width=15cm}}
\caption
{Imaginary part of electronic self-energy contributions versus frequency
for $\lambda=1.0$.
The solid line refers to the diagram (a) and the dashed line to the diagram (b) of
Fig.1.}
\label{Migdal2}
\end{figure}
\newpage

\begin{figure}
\centerline{\psfig{figure=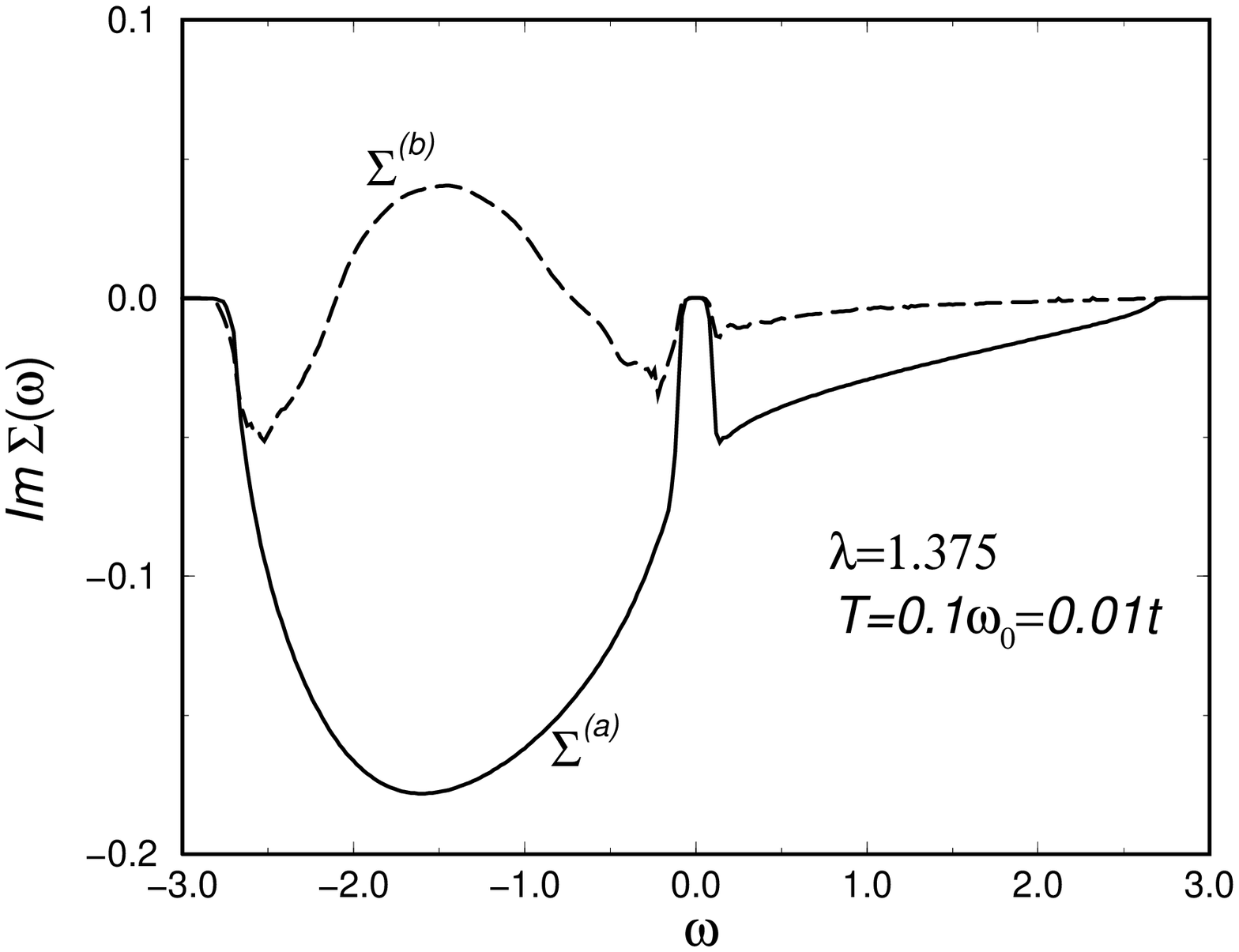,width=15cm}}

\caption
{Imaginary part of electronic self-energy contributions versus frequency
for $\lambda=1.375$.
The solid line refers to the diagram (a) and the dashed line to the diagram (b) of
Fig.1.}
\label{Migdal3}
\end{figure}
\newpage

\begin{figure}
\centerline{\psfig{figure=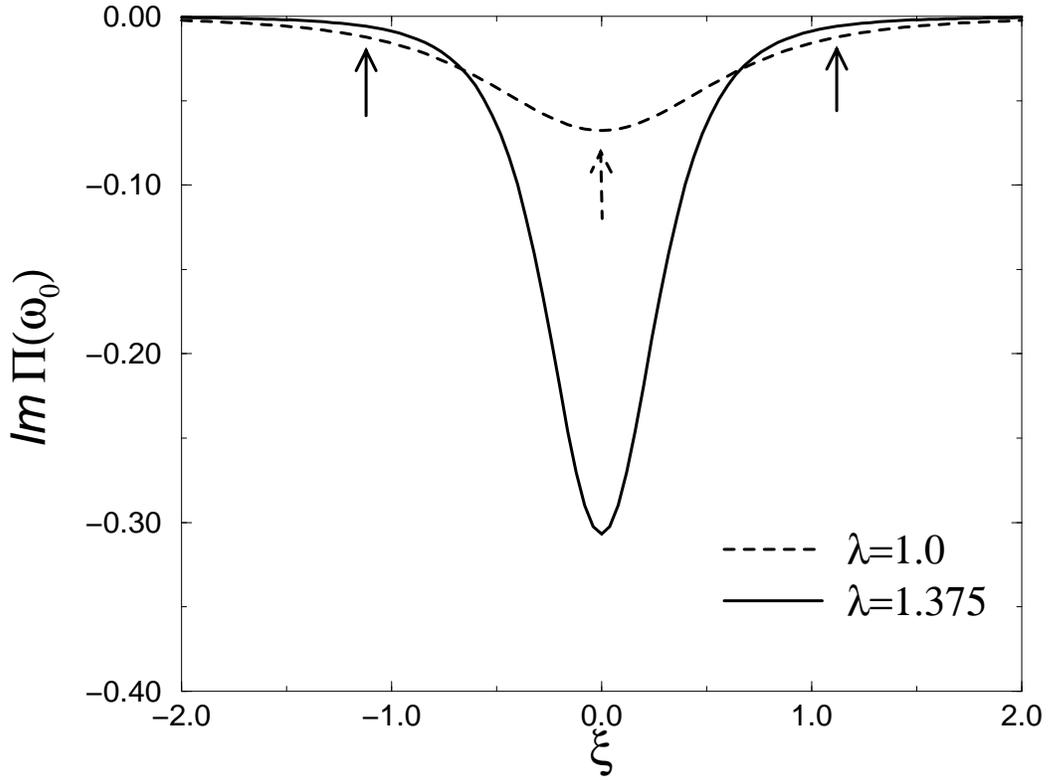,width=15cm}}
\caption
{Imaginary part of the phonon self-energy $\Pi(\omega_{0})$
as a function of the dimensionless field 
$\xi= \tilde{\phi}\, \omega_{0}/t^{1/2}$ for $T=0.1\omega_{0}=0.01t$. The arrows show the correspondent
position of the minima of the local effective potential.}
\label{Buble}
\end{figure}
\newpage

\begin{figure}
\centerline{\psfig{figure=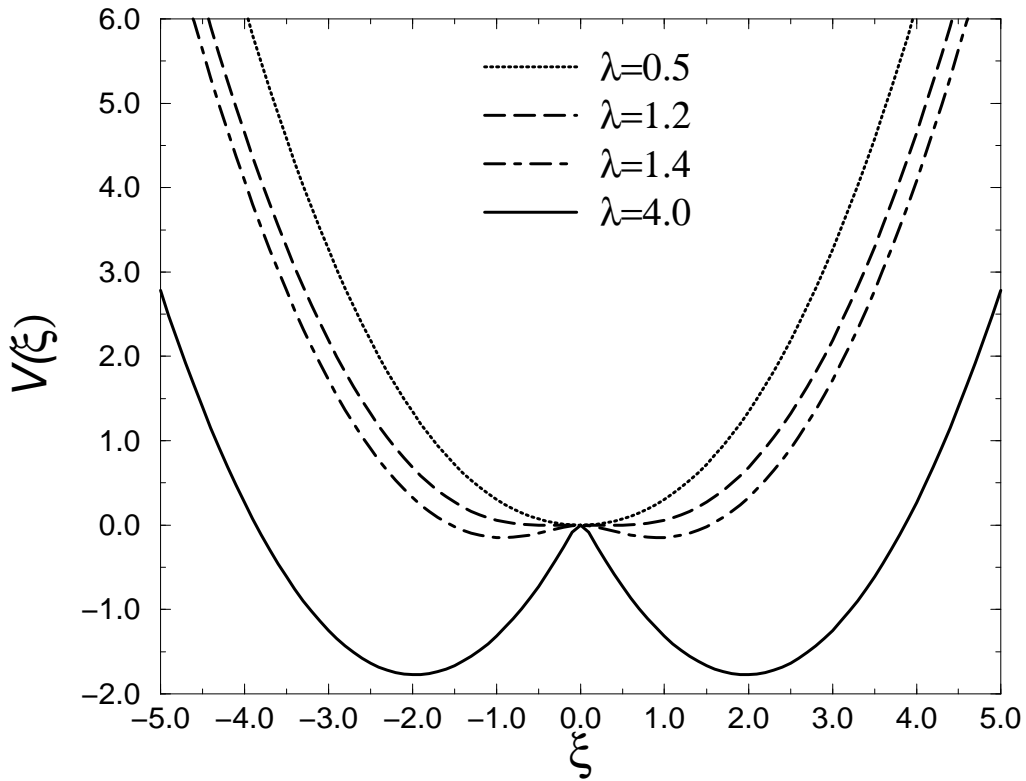,width=15cm}}
\caption
{Effective potential $V$ of the atom  versus the dimensionless
phonon field $\xi=\tilde{\phi}\, \omega_{0}/ t^{1/2}$ 
for $T=0.1\omega_{0}=0.01t$ and different $\lambda$'s. }
\label{Pot.}
\end{figure}
\newpage

\begin{figure}
\centerline{\psfig{figure=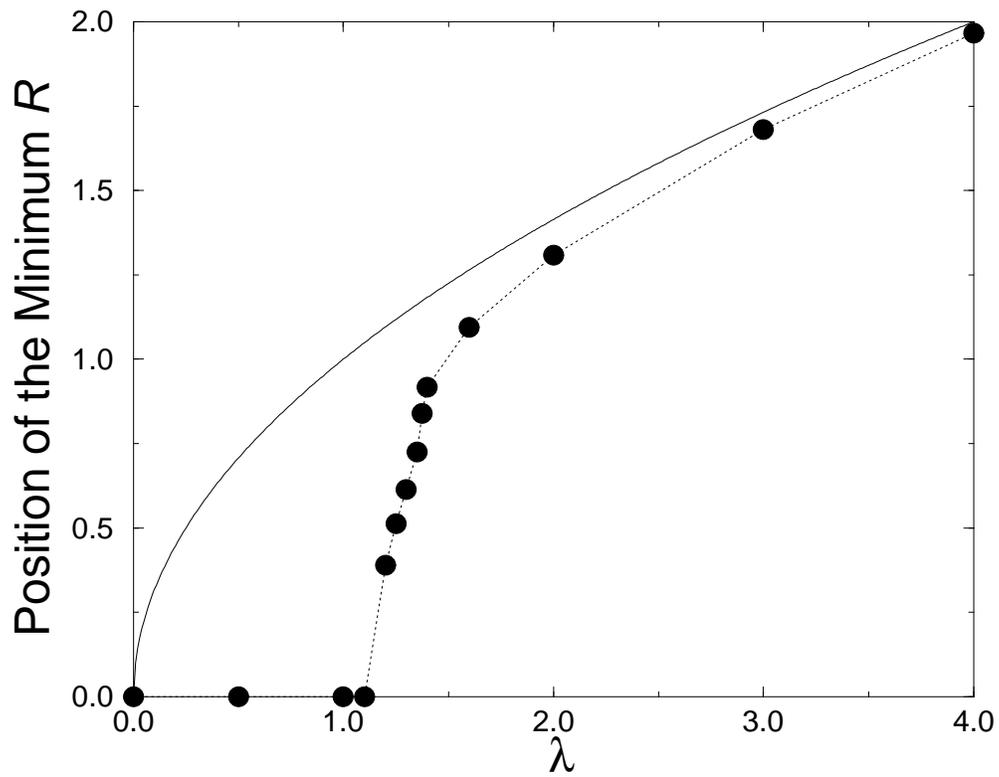,width=15cm}}
\caption
{Dotted line with circles: Position of the potential minimum R 
as a function of $\lambda$, for $T=0.1\omega_{0}=0.01t$. Solid line:
Asymptotic $\lambda^{1/2}$ dependence for vanishing hopping.}
\label{R} 
\end{figure}
\newpage

\begin{figure}
\centerline{\psfig{figure=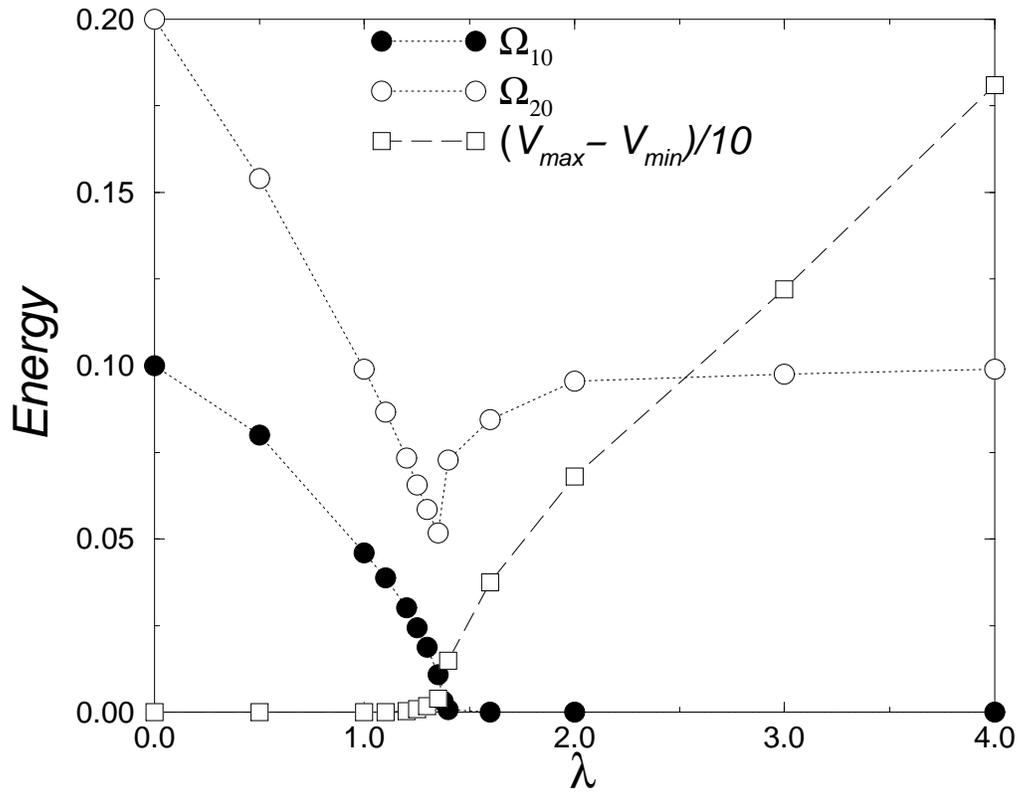,width=15cm}}
\caption
{Potential barrier $V_{max} -V_{min}$ (squares) and energy differences between
the ground state and the first (filled circles) and the second (empty circles)
excited states of the effective potential as a function of $\lambda$
for $T=0.1\omega_{0}=0.01t$. }
\label{Corr.el-ph}
\end{figure}
\newpage

\begin{figure}
\centerline{\psfig{figure=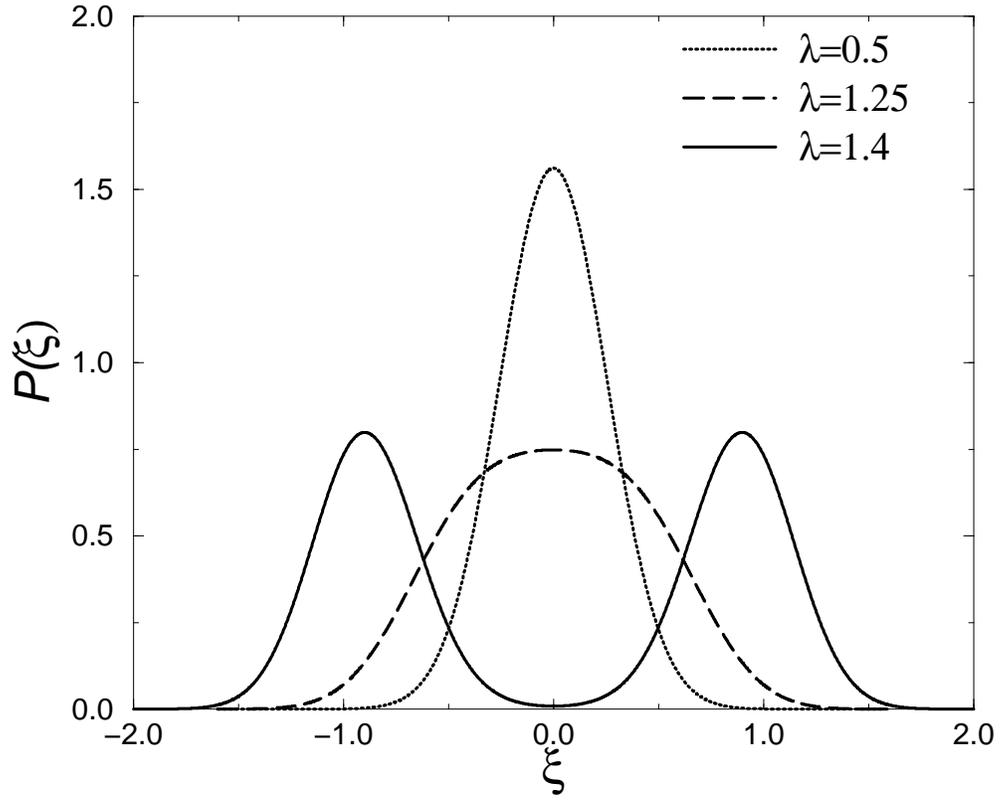,width=15cm}}
\caption
{Probability distribution $P$ of the atom as a function of
$\xi= \tilde{\phi}\, \omega_{0}/t^{1/2}$ for $T=0.1\omega_{0}=0.01t$ and three
different values of $\lambda$. }
\label{Prob.}
\end{figure}
\newpage

\begin{figure}
\centerline{\psfig{figure=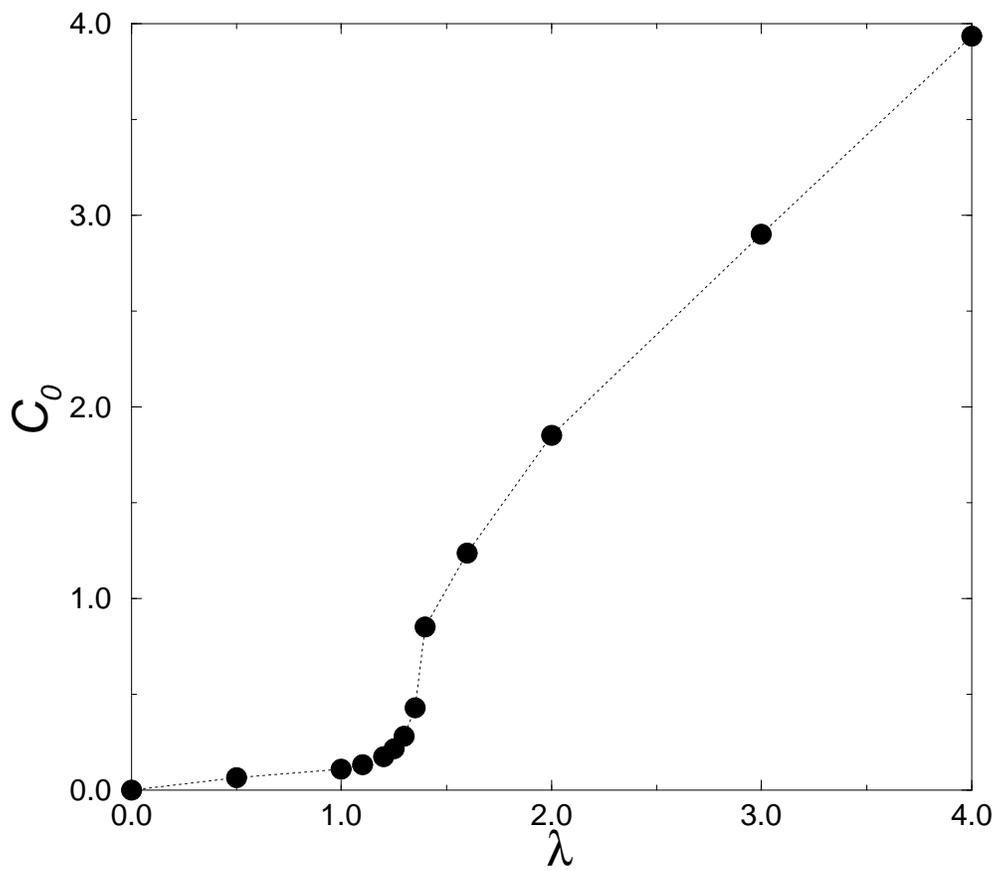,width=15cm}}
\caption
{Static electron-phonon correlation function 
$C_{0}=-g/t \left<\tilde \phi\,n\right>$
versus $\lambda$ for $T=0.1\omega_{0}=0.01t$.}
\end{figure}
\newpage

\begin{figure}
\centerline{\psfig{figure=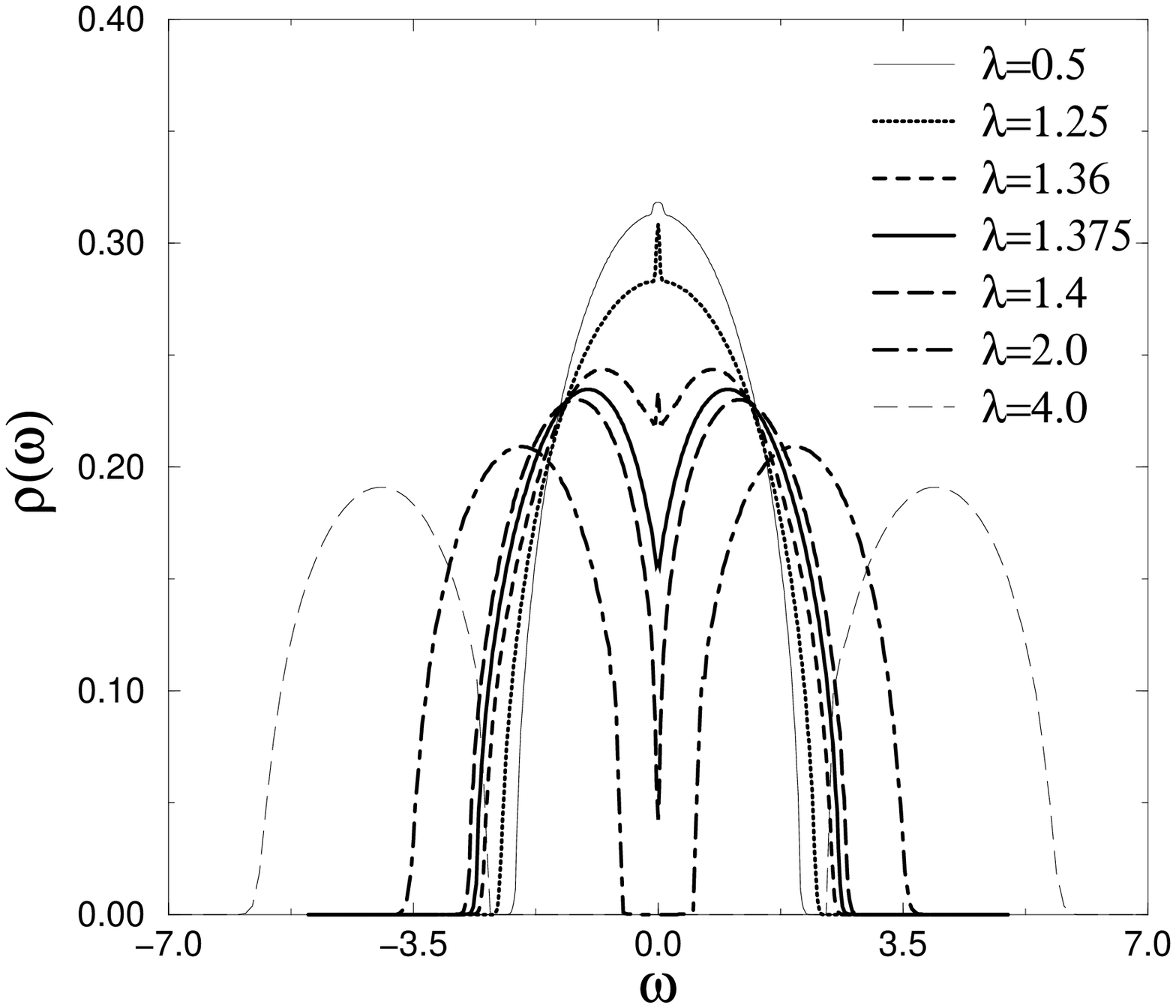 ,width=15cm}}
\caption
{Electronic density of states $\rho$ versus frequency 
for  $T=0.1\omega_{0}=0.01t$
and different $\lambda$'s.}
\label{Rho}
\end{figure}
\newpage
\begin{figure}
\centerline{\psfig{figure=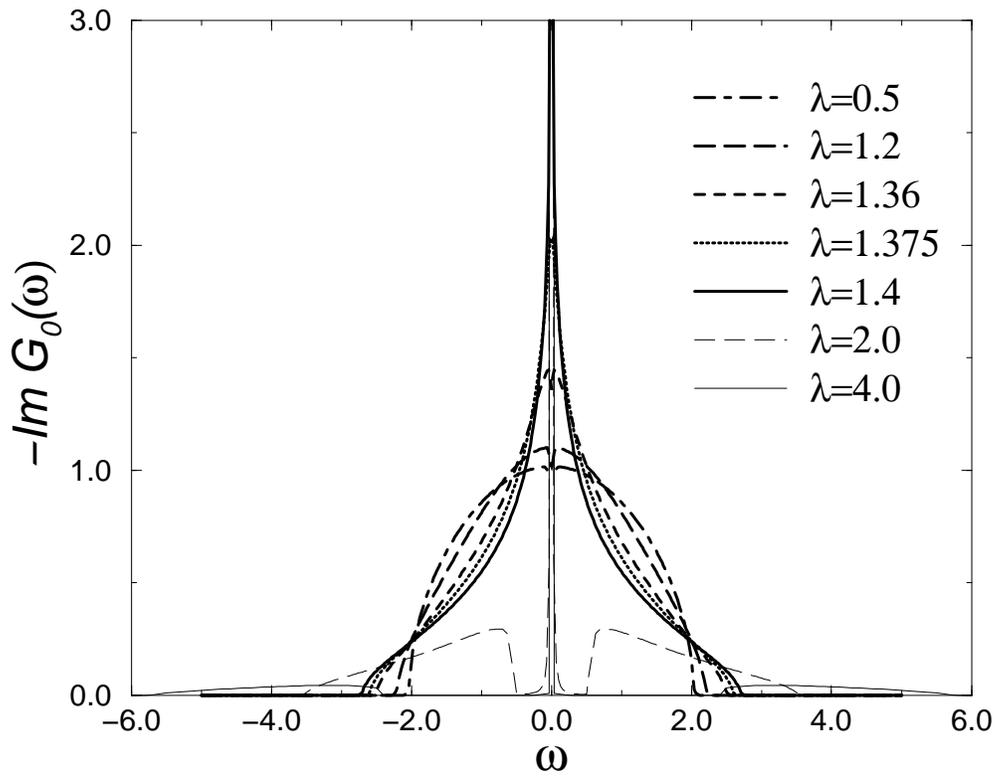 ,width=15cm}}
\caption
{Imaginary part of the "bare" electronic Green's function $G_{0}$ versus 
frequency for $T=0.1\omega_{0}=0.01t$ and different $\lambda$'s.}
\label{Go}
\end{figure}
\newpage
\mbox{}\mbox{}\mbox{}
\begin{figure}
\centerline{\psfig{figure=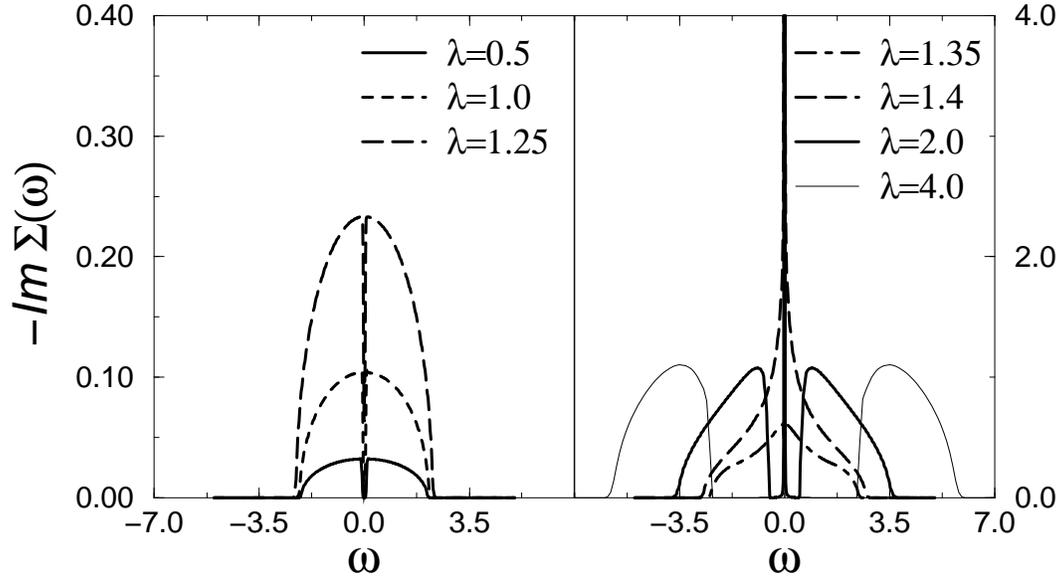 ,width=15cm}}
\caption
{Imaginary part of the electronic self-energy $\Sigma$ versus frequency 
for $T=0.1\omega_{0}=0.01t$. 
The chosen values of $\lambda$ extend from the weak- (a) to the
strong-coupling (b) regime.}
\label{Sigma}
\end{figure}
\newpage
\begin{figure}
\centerline{\psfig{figure=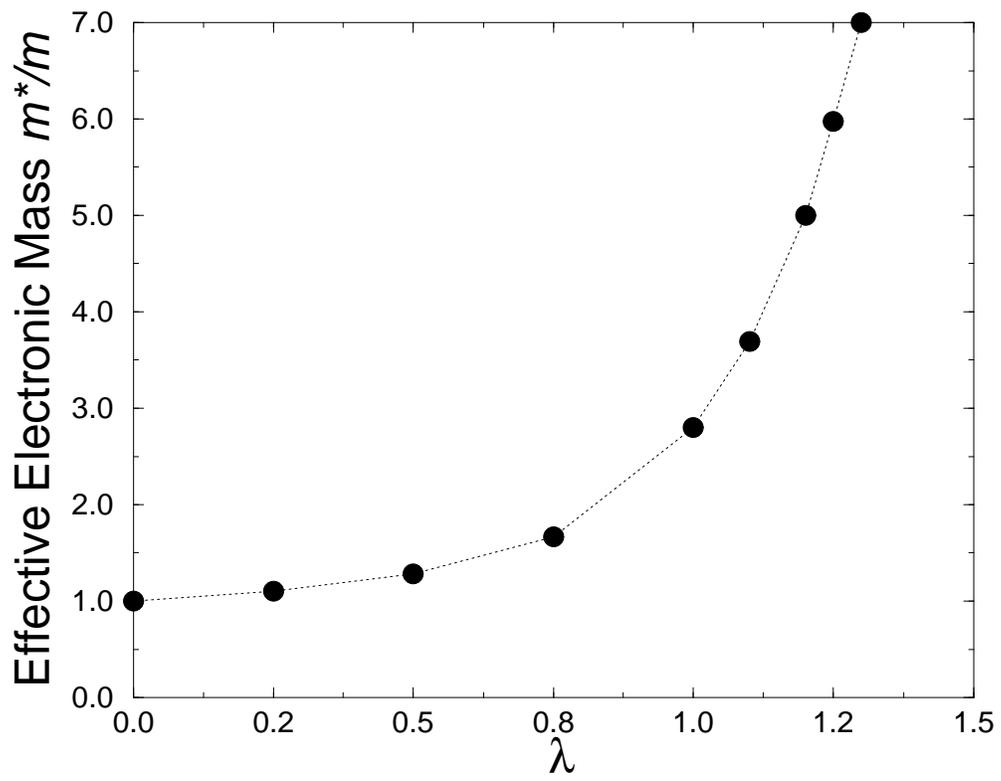 ,width=15cm}}
\caption
{Ratio of the effective and bare electronic mass versus
 $\lambda$ for $T=0.1\omega_{0}=0.01t$. }
\label{M*}
\end{figure}
\newpage
\begin{figure}
\centerline{\psfig{figure=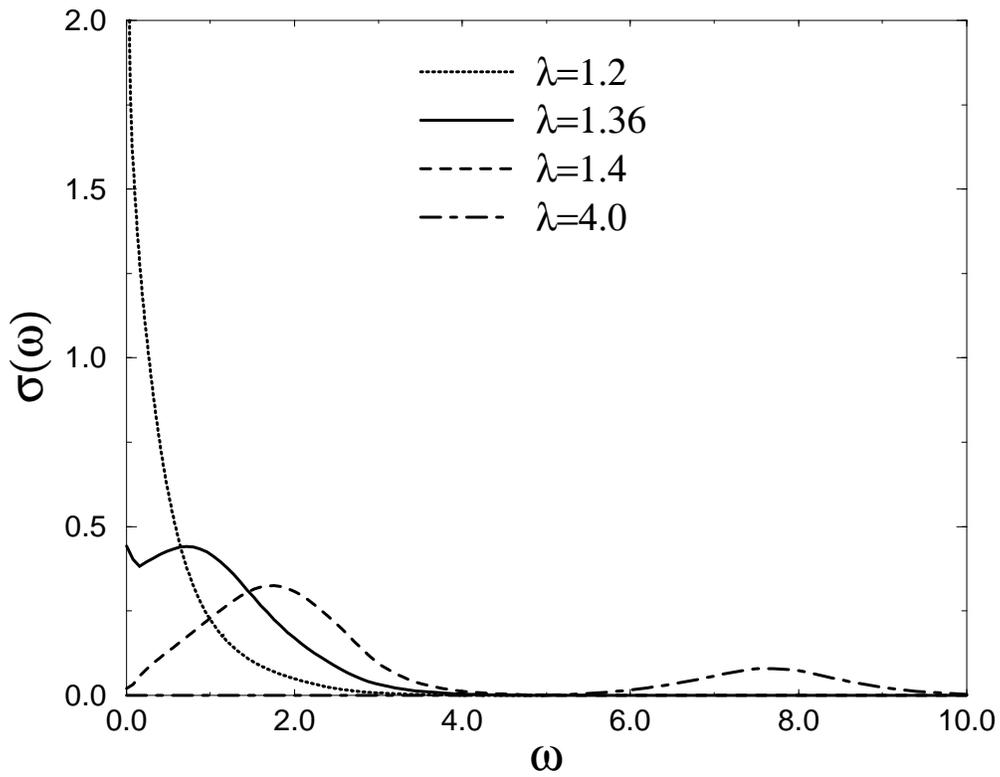 ,width=15cm}}
\caption
{Optical conductivity $\sigma$ in arbitrary units as a 
function of the frequency for $T=0.1\omega_{0}=0.01t$ and 
different $\lambda$'s.}
\label{Conduct.}
\end{figure}
\newpage
\begin{figure}
\centerline{\psfig{figure=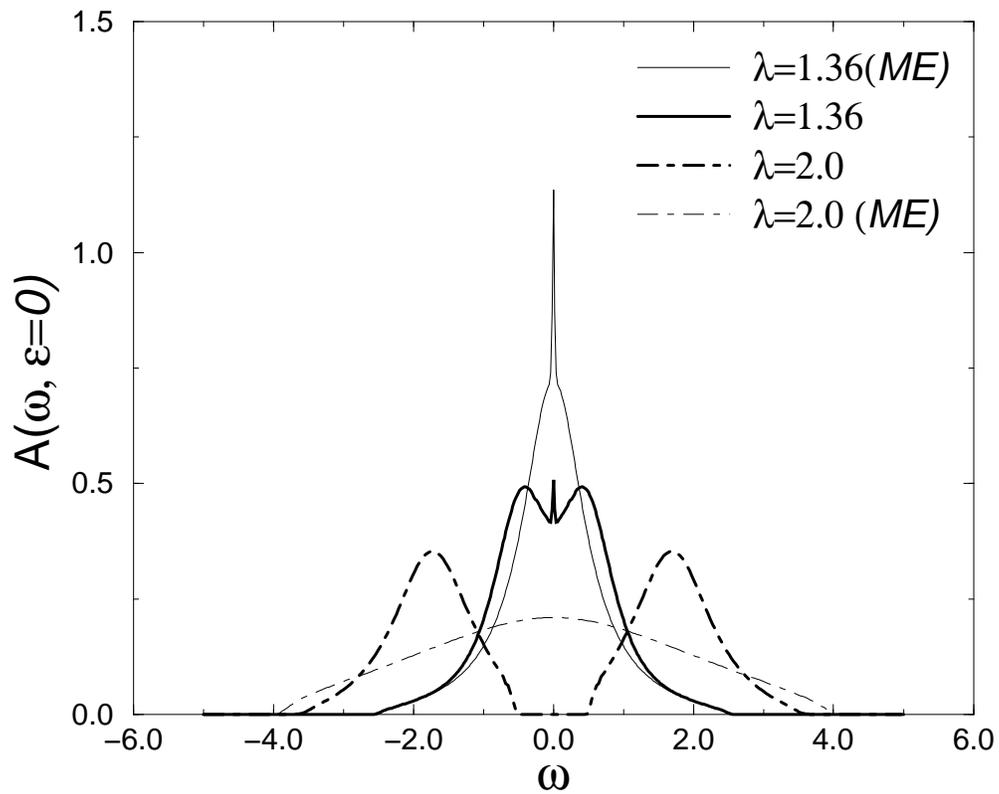 ,width=15cm}}
\caption
{One-particle spectral function $A$ at the Fermi surface as a 
function of the frequency for $T=0.1\omega_{0}=0.01t$. The thin solid and
dot-dashed lines are calculated using the Migdal-Eliashberg theory, the
corresponding thick lines are the present results. }
\label{ME}
\end{figure}
\newpage
\begin{figure}
\centerline{\psfig{figure=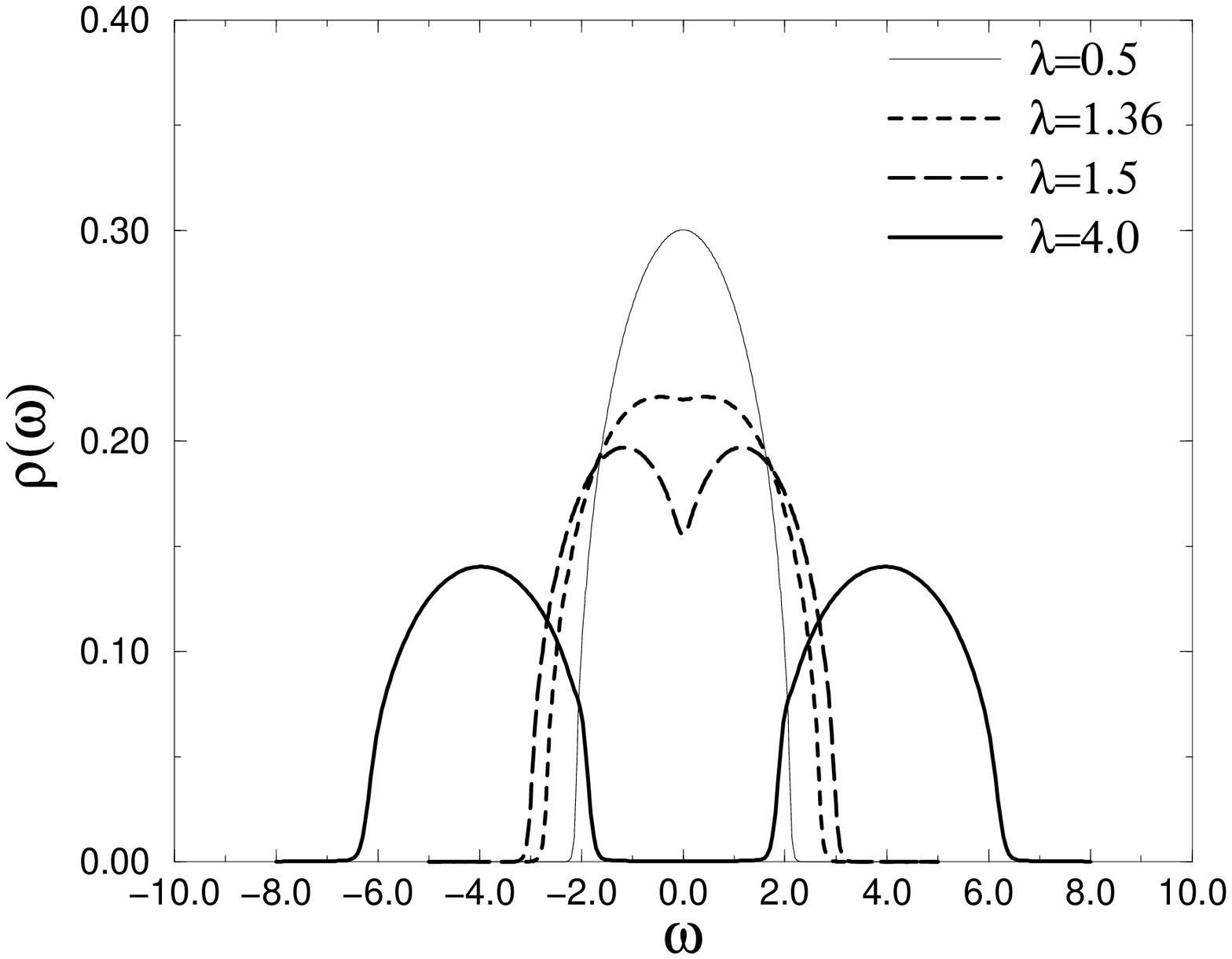 ,width=15cm}}
\caption
{Electronic density of states $\rho$ versus frequency for
$T=2.0\omega_{0}=0.2t$ and different $\lambda$'s .}
\label{Rho_HT}
\end{figure}

\begin{figure}
\centerline{\psfig{figure=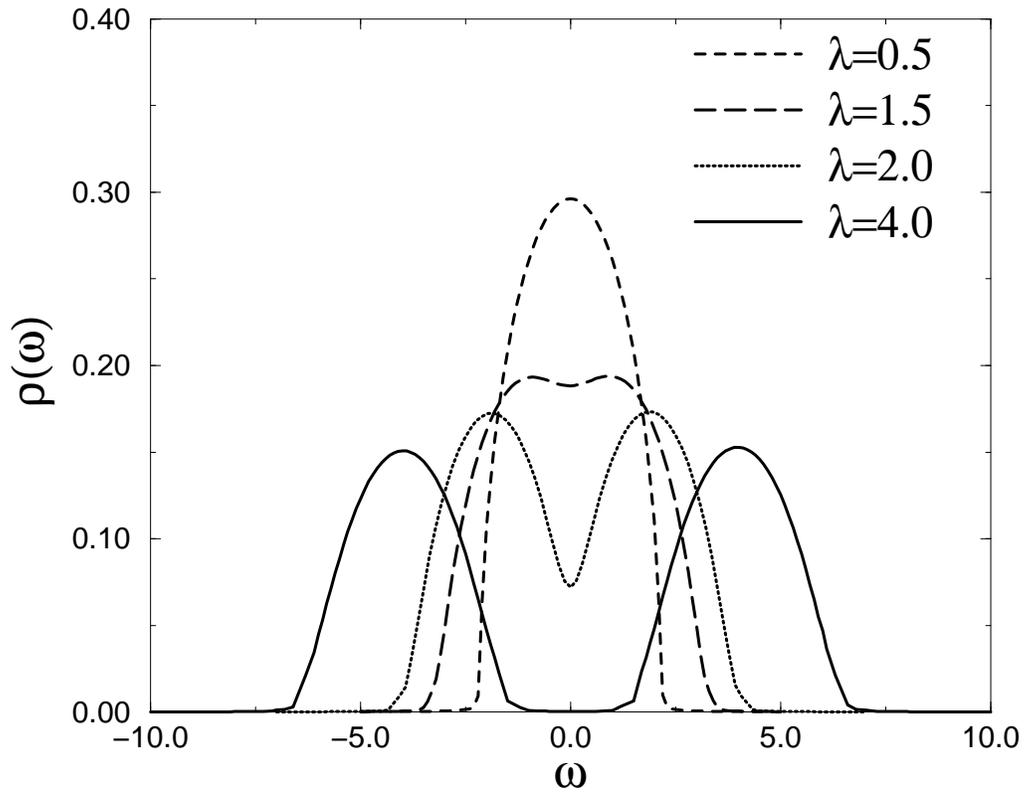,width=15cm}}
\caption
{Electronic density of state $\rho$ versus frequency for $T=2.0\omega_{0}=0.2t$ 
and different $\lambda$'s in the limit of classical, infinite heavy atoms.}
\label{Rho_HTcl}
\end{figure}
\newpage
    
\end{document}